\documentclass[12pt] {article}
\usepackage{amstex}
\usepackage{epsfig}

\def\CC{\hbox{\it l\hskip -5.5pt C\/}}
\def\RR{\hbox{\it I\hskip -2.pt R }}
\def\MC{\hbox{\it I\hskip -2.pt M \hskip -7 pt I \hskip - 3.pt \CC}_n}
\newcommand{\nc}{\newcommand}
\nc{\beq}{\begin{equation}}
\nc{\eeq}{\end{equation}}
\nc{\beqa}{\begin{eqnarray}}
\nc{\eeqa}{\end{eqnarray}}
\def\agt{
\mathrel{\raise.3ex\hbox{$>$}\mkern-14mu\lower0.6ex\hbox{$\sim$}}
}
\def\alt{
\mathrel{\raise.3ex\hbox{$<$}\mkern-14mu\lower0.6ex\hbox{$\sim$}}
}
\begin {document}
\bibliographystyle{unsrt}    

\hspace{13cm}PM-99/25

\vspace{2cm}

\Large
{\centerline{\bf Hidden symmetries in quantum field theories}
\vspace{3mm}
{\centerline{\bf from extended complex numbers}}
\vspace{2mm}

\large

\vspace{1.5cm}

\centerline {\bf Pascal~Baseilhac}

\small\normalsize
\vspace{3mm}

\centerline {Laboratoire de Physique Math\'ematique et Th\'eorique}

\vspace{3mm}

\centerline {UMR CNRS 5825, Universit\'e Montpellier II}

\vspace{3mm}

\centerline {Place E.~Bataillon, 34095 Montpellier, France}

\vspace{7mm}

\vspace{1cm}

\begin{abstract}
The 2-dimensional space-time sine-Gordon field theory is extended algebraically within the $n-$dimensional space of extended complex numbers. This field theory is constructed in terms of an adapted extension of standard vertex operators. A whole set of non-local conserved charges is constructed and studied in this framework. Thereby, an algebraic non-perturbative description is possible for this $n-1$ parameters family of quantum field theories. Known results are obtained for specific values of the parameters, especially in relation to affine Toda field theories. Different (dual)-models can then be described in this formalism.    
\end{abstract}


\newpage

\section{Introduction}\label{Introduction}

In most domain of physics, probably related to the bilinear aspects of fundamental objects such as quadratic metric, commutators, anticommutators, etc...algebras induced by {\it bilinear} relations have a special status. Clifford algebras of polynomials of degree higher than two \cite{cp,cp2} have been constructed thirty years ago by mathematicians. Within this family, the natural algebraic extensions of the Clifford and Grassmann algebras \cite{line,gca,gca2}, lies the extensions of complex numbers \cite{mc1,mc2}. It leads to the corresponding extensions of trigonometric functions, dubbed multisine functions.

It is thus natural to study some possible extensions of a field theory described in terms of  $\mathbb C$-valued vertex operators in its ``simplest'' (apparently) famous form, the sine-Gordon  (SG) model, known for its integrability and early established boson-fermion duality property \cite{Col,Zamo,Kla}. For instance, Toda and affine Toda field theories based on simply or non-simply laced algebras are the natural simple Lie group extension of the $SU(2)$ and $\hat{SU}(2)$ Toda (TFT) and affine Toda field theories (ATFT) \cite{brad}, {\it e.g.} Liouville and SG models. However, vertex operators description of these Lie algebras extensions remains $\mathbb C$-dependent.

Independently, certain quantum field theories (QFTs) for which non-perturbative computations are known admit an infinite number of conservation laws \cite{pohl,lu1}. On the one hand exact $S$-matrices in two dimensional models can be obtained from the Yang-Baxter relations \cite{zam,dev} which are a direct consequence of the conservation of higher powers of the momentum. On the other hand, in lower-dimensional quantum field theory, some of the postulates of the Coleman-Mandula theorem \cite{0} can be relaxed in a non-trivial way : non-local conserved charges \cite{lu1} may appear that generate symmetries, characterized by non-trivial equal-time commutation, or braiding relations \cite{15}. The first non-trivial one alone fixes almost completly the on-shell dynamics \cite{lu1,abda}. The action of the generator of the symmetry group on a multiparticle state is identified to the non-trivial comultiplication of the charges. A well-known example corresponds to the quantum group symmetry algebra $U_q({\hat {sl_2}})$ associated to the sine-Gordon model \cite{16}. This symmetry is then sufficiently restrictive to allow a non-perturbative solution of the theory : conservation of a non-local charge imposes strong constraints on the $S$ matrix. In addition, the whole set of non-local conserved charges constitutes an infinite dimensional non-abelian symmetry of Yang-Baxter equations. In particular, one finds the quantum group invariance as a limiting case \cite{17}. Consequently, non-local conserved charges are very powerfull objects appearing in many other field theories \cite{18,Del}. 
 
More recently, the interest in different integrable deformations of ATFT came from their integrability and duality properties \cite{Gris,Fateev1} which can be used to pave the way to the understanding of the electric-magnetic duality in four-dimensional gauge theories, conjectured in \cite{elec-mag duality1} and developped in \cite{elec-mag duality2}. Thereby non-perturbative analysis of the spectrum and phase structure in supersymmetric Yang-Mills theory becomes possible.

In an earlier study \cite{multisine} and elsewhere \cite{multisine2} we defined the multisine-Gordon (MSG) models, a wide variety of $n-1$ parameters $(\{\alpha_a\},\{\beta_a\})$ QFTs, for two different types of representations associated to the multicomplex algebra, called ``minimal'' and ``non-minimal''. For specific values of the multicomplex dimension and of the parameters, we prooved the integrability of different cases (for ``minimal '' and ``non-minimal'' representations) by constructing {\it local} conserved charges. In particular, we presented their connections with different types of integrable deformations of the non-linear $\sigma$ model, parafermionic sine-Gordon model, etc...

The purpose of this paper is to investigate {\it non-local} charges algebras which may appear in those quantum field theories in $1+1$ dimensions. Using the multicomplex formalism, we extend to these family of models the framework developped in the past for the sine-Gordon theory and more generally for affine Toda field theories \cite{16}. This point of view provides a new insight on apparently disconnected models.

In section 2, the mathematical formalism associated to the multicomplex algebra is introduced. Algebras and ``non-minimal'' representations properties (described in \cite{multisine2}) are briefly recalled.  The natural multicomplex extension of vertex operators is defined, namely MC-vertex operators. The corresponding OPE rules are then given.

Section 3 opens with the formalism developped by Zamolodchikov \cite{20} and related to the construction of conserved charges. The approach to be applied to the MSG model is then defined. A set of non-local {\it multicomplex} charges results if several constraints on parameters are verified. The {\it multicomplex} algebra associated to these non-local charges is explicitely found and characterized by a quantum deformation and the existence of a topological term. The structure simplifies when restricting the space of parameters to  ($\{\alpha_a\}\in \mathbb{R}^{n/2-1}$, $\{\beta_a\}\in i\mathbb{R}^{n/2-2}$) for a specific (``minimal'' or ``non-minimal'') representation of multicomplex algebra in even $n$ dimension. A whole set of non-local conserved charges of the MSG model is obtained. Performing a restriction of these results leads to informations about hidden symmetries for odd $n$ dimensions.

Some specific examples are studied in section 4 for a particular representation of multicomplex algebra. For $n=3$, it is shown that, for different values of parameters $\alpha_0$, $\beta_0$ which identify $MSG_{(3|3)}$ and $MSG_{(3|4)}$ to $A_2^{(1)}$, $B_2^{(2)}=D_3^{(2)}$ and $C_2^{(1)}$ ATFTs, our formula for the non-local charges algebra are in perfect agreement with known results ($U_q(A_2^{(1)})$, $U_q(C_2^{(1)})$ and $U_q(B_2^{(2)})$ respectively). Similarly, for $n=4$, different parameters values give the connection with known results. A first example is the restriction identified to two sine-Gordon models for which the non-local charges algebra decouples trivially in a product of two $U_q({\hat {sl_2}})$ quantum envelopping algebras. An other choice of parameters connects the MSG with $A_3^{(1)}$ ATFT and identifies the corresponding non-local charges algebra to $U_q(A_3^{(1)})$ quantum envelopping algebra. Other restrictions are also considered.

Some conclusions and perspectives are drawn in section 5.

\section{Multicomplex extension of vertex operators}
Here some basic properties of the multicomplex (MC) numbers are collected that are relevant for the sequel, for more details one can see \cite{mc1,mc2}. The set of MC-numbers is  generated by one element $e$ 
which satisfies $e^n=-1$ \cite{mc1} 
($\MC=\left\{x= \sum \limits_{i=0}^{n-1} x_i e^i,
         e^n=-1, x_i \in \RR\right\}$),
and constitute a $n-$dimensional commutative algebra. It is worth stressing 
that most of the results of usual complex mumbers analysis remains true for MC-numbers, were it be for algebraic \cite{mc1} or analytic properties \cite{mc2}.

Introducing $x^{(l)}=\sum\limits_{i=0}^{n-1} x_i q^{il}e^i$,
the conjugate of order $l$ of $x^{(0)}$, ($q=\exp(2 i \pi/n)$), one may define the pseudo-norm \cite{mc1,mc2}\ $\|x\|^n= \prod \limits_{l=0}^{n-1}x^{(l)}$ and, if $\|x\| \neq 0$, $x$ can be written in an exponential form
\beqa\label{1}
x=\sum \limits_{i=0}^{n-1}x_i e^i  
= \rho  \exp{\left(\sum \limits_{i=1}^{n-1} \Theta_i e^i\right)}, \ \  \ \mbox{with} \ \ \|x\|=\rho.
\eeqa
Of course, due to the multivaluedness of the logarithm function the $\Theta$'s are not unique. Then, it becomes  straightforward to define the extension of the usual trigonometric functions \cite{mc1}
$x=\rho \exp{\left(\sum \limits_{i=1}^{n-1} \Theta_ie^i \right)}=
\rho \sum \limits_{i=0}^{n-1} \mathrm{mus}_i(\Theta) e^i$.
These functions have properties analogous to the usual sine and cosine functions
\cite{mc1}. However, the parametrisation in terms of $\Theta$ given in \cite{mc1,mc2} is inconvenient to derive explicit formulas for the multisine functions. From the properties of $e$ in its matrix representation (see \cite{mc1}) it is easy to see that $e^i+e^{n-i}$ (resp. $e^i-e^{n-i}$) are represented by pure imaginary (real) matrices, and, hence among the $n-1$ directions in $\sum \limits_{i=1}^{n-1} \Theta_i e^i$ one can extract $E(n/2)$ compact directions and $n-1-E(n/2)$ non-compact ones ($E(x)$ denotes the integer part of $x$). For instance, performing the change of basis, we have for even $n$ $\sum \limits_{i=1}^{n-1} \Theta_i e^i = \sum \limits_{a=0}^{n/2-1} 
\Big[\phi_a P_a + \varphi_a (-P_a^2) \Big]$, with
$P_a$ ($a=0, \cdots, n/2-1$) given by $P_a= {2 \over n} \sum \limits_{j=0}^{n-1} \sin\Big[(2a+1) j {\pi \over n}\Big]e^j$. For odd $n$, the generators are slightly different, however in the following sections we shall not need their explicit expressions. 

The multisine functions related to the lowest $n \times n$ matrix 
representation are in this sense the ``minimal'' ones that can be defined. However, using higher dimensional matrix realizations of $\MC$ other series for the multisine functions are also allowed \cite{multisine2}. For instance, if the natural embedding, $ \MC \subset \hbox{\it I\hskip -2.pt M \hskip -7 pt I \hskip - 3.pt \CC}_m$ is used, $n<m$ with any $m$ when $n$ is odd, and even $m$ when $n$ is even. Denoting $e_m=(-1)^m\exp(\varepsilon_m)$, with $\varepsilon_m= \sum \limits_{i=1}^{n-1}\varepsilon_{m,i} e^i$ the generators of $\hbox{\it I\hskip -2.pt M \hskip -7 pt I \hskip - 3.pt \CC}_m$, one may define $e_{(n|m)}=(-1)^n \exp({m \over n} \times \varepsilon_m)
\equiv \left(e_m\right)^{{m \over n}}$. It is easy to see that $e_{(n|m)}$ generates $\MC$. Furthermore, defining the multisine functions built from the $e_{(n|m)}$, with the $e_{(n|m)}$ understood as an element of $\hbox{\it I\hskip -2.pt M  \hskip -7 pt I \hskip - 3.pt \CC}_m$, it is clear that the multisine functions so obtained  satisfy an equation of degree $m$ instead of $n$. In other words, the multisine of order $n$ can be constructed from linear combinations of the multisine of order $m$. It is not difficult to derive the relation fulfilled by the multisine functions obtained in the \ $m \times m$\  matrix realization. Indeed, one can immediately see that with such an embedding $e_{(n|m)}$ is represented by a diagonal matrix with $(q^{1/2},q^{-1/2})$  repeated $m_0$ times, $(q^{3/2},q^{-3/2})$ $m_1$ times, etc... The $m_i$'s $\ge 1$ depend of course of $m$ and $n$\,\footnote{More generally, one can consider other higher dimensional representations, which cannot always be  understood along the lines of the embedding, {\it i.e.}  when $m_0 \ge 1$, $m_1 \ge 1$, etc... are arbitrary numbers.}. The relation obeyed by the multisine functions in the ``non-minimal''case can then be easely obtained. Of course, if all the $m_i$'s are equal to the same single value $p$ we obtain nothing new. This just comes from the embedding
$\hbox{\it I\hskip -2.pt M \hskip -7 pt I \hskip - 3.pt \CC}_n \subset
\hbox{\it I\hskip -2.pt M \hskip -7 pt I \hskip - 3.pt \CC}_{pn}$.
As a consequence,  $\CC \subset 
\hbox{\it I\hskip -2.pt M \hskip -7 pt I \hskip - 3.pt \CC}_{2n}$
leads always to the usual sine and cosine functions. Finally, the explicit formulas for the multisine functions obtained from the
$(m_0+m_1+\cdots )\times (m_0+m_1+\cdots )$ dimensional
matrix representation are the same as those obtained in the ``minimal case''. The only difference appears in the constraint among the non-compact variables (resulting from the unity condition on the determinant value) \cite{multisine,multisine2}.

The standard vertex operators are defined by the expressions \ \ $J_\alpha = \exp(i\alpha \phi(z))$\ \ (holomorphic part) and \ \ $ \overline{J}_\alpha = \exp(i\alpha \overline{\phi(z)})$\ \ (anti-holomorphic part) with $\alpha\in \mathbb{C}$. In analogy the following writtings are introduced :

\vspace{2mm}

{\bf Definition 1} : {\it The $n$-multicomplex extension of the vertex operators, namely MC-vertex operators\,\footnote{Vertex operators are always implicitely normal-ordered.} is defined as :}
\beqa
\ \ \ \ \ \ \ J_{\eta^{(k)}} &=& J^{(k)} = \exp\big(\eta^{(k)}.\phi(z) \big)\ \ \ \  \ \ \ \mbox{(holomorphic part)},\label{mvertex} \\
\overline{J}_{\eta^{(k)}} &=& \overline{J}^{(k)} = \exp\big(\eta^{(k)}.\overline{\phi(z)}\big) \ \ \ \ \ \mbox{(anti-holomorphic part)},\nonumber
\eeqa
where $k=0,...,n-1$ and ($\phi(z),\overline{\phi(z)}$) are holomorphic and anti-holomorphic part of a $n$ component vector field $\Phi(z,\overline{z})$\,\footnote{Strictly speaking, we cannot write $\Phi=\phi+{\overline \phi}$ because the zero mode would be duplicated in this process. However, the zero mode is absorbed in the definition of $\phi$. Then, the full vertex operator decomposes into a product of left and right chiral vertex operators.}. The Euclidian light-cone coordinates are $(z,{\overline z})$ as $z=i(x+t)/2$ and ${\overline z}=i(t-x)/2$. By convention, we have $d^2z=idzd{\overline z}=-idxdt/2$. The expression \ \ $\eta^{(k)}.\phi(z)$ \ \ (resp. \ \ $\eta^{(k)}.\overline{\phi(z)}$ ) is defined in the next paragraphs as it depends on the parity of $n$. 

\vspace{7mm}  
\centerline{\bf Even multicomplex extension}
\vspace{3mm}

If the dimension $n$ of the multicomplex space is even, the holomorphic part (and similarly for the anti-holomorphic part) \cite{multisine,multisine2} are defined as : 
\beqa
\eta^{(k)}.\phi(z)= \sum_{a=0}^{\frac{n}{2}-1} \Big[ \alpha_a P_{a+k}\phi_a(z) + \beta_a (-P^2_{a+k})\varphi_a(z) \Big],\label{d1}
\eeqa
\beqa
\mbox{where}\ \ \ \eta^{(k)}&=& [\alpha_0 P_{k},...,\alpha_{\frac{n}{2}-1} P_{\frac{n}{2}-1+k }; -\beta_0 P^2_{k} ,...,-\beta_{\frac{n}{2}-1}P^2_{\frac{n}{2}-1+k}],\label{etak0}\\
 \phi(z)&=& [\phi_0(z),...,\phi_{\frac{n}{2}-1}(z); \varphi_0(z),...,\varphi_{\frac{n}{2}-1}(z)]\nonumber
\eeqa
are $n$ multicomplex components vectors, with $(\{\alpha_a\}, \{\beta_a\})$ the extensions of the complex parameter $\alpha$ in $J_\alpha$. The $n$-multicomplex algebra of the generators $P_{a}$ is defined, for $(a,b,k)\in[0,...,n/2-1]$, as :
\vspace{-0.2cm}
\beqa
P_{a}P_{b}&=&\delta_{a,b}P^2_a \ \ \ \ \  \mbox{with} \ \ \ \ \delta_{a,b}\ \ \ \ \mbox{the kr\" onecker symbol},\label{pa}\\
P_{a+n/2}&=& -P_{a},\ \ \ P_{a+n}= P_{a}, \nonumber\\
\ \ \mbox{and} &&\ \ \ \sum_{a=0}^{n/2-1}-P^2_{a}= 1.\nonumber       
\eeqa
For example, one faithful representation $\pi$ is given by $(n\times n)$ dimensional diagonal matrices:
\beqa
&&\pi\big[P_{a}\big]= Diag(0,...,0,i,0,...,0,-i,0,...,0)\ \  \  \ \mbox{for} \ \ a\in[0,...,n/2-1],\label{rep} \\
&& \mbox{with} \  i(\mbox{resp.} -i)\ \mbox{is in the}\ a\ (\mbox{resp.\ } n-1-a)\ \mbox{position},\nonumber \\
\nonumber\\
&&e.g.\  {(\pi\big[P_{a+k}\big])}_{jj}= (\pi\big[{P_{a+k}}^{\dagger}\big])_{n-1-j,n-1-j}\ \ \mbox{where $ ^{\dagger}$ denotes the ordinary hermitian conjugate},\nonumber\\
    && \ \mbox{for}\ \ \  (a,k)\in[0,...,n/2-1]\ \mbox{and}\ j\in[0,...,n-1].\nonumber
\eeqa
As indicated in \cite{multisine,multisine2}, to obtain unimodular $n$-multicomplex numbers, it is necessary to have a constraint over the $n/2$ non-compact fields $\varphi_a$ which restricts the independent degrees of freedom of the theory from $n$ to $n-1$ :
\beqa\label{cont}
m_0\beta_0\varphi_0 +...+ m_{\frac{n}{2}-1}\beta_{\frac{n}{2}-1}\varphi_{\frac{n}{2}-1}=0.
\eeqa
Then, with the constraint (\ref{cont}), expression (\ref{d1}) is modified as :  
\beqa
\eta^{(k)}.\phi(z)= \sum_{a=0}^{\frac{n}{2}-1} \Big[ \alpha_a P_{a+k}\phi_a(z)\Big] + \sum_{a=0}^{\frac{n}{2}-2} \Big[\beta_a Q_{a;k}\varphi_a(z) \Big]\label{d2}
\eeqa
with :
\beqa
Q_{a;k}&=&-P_{a+k}^2+\frac{m_a}{m_{n/2-1}}P^2_{n/2-1+k }\ \ \  \mbox{for}\  a\in[0,...,n/2-2]\  \mbox{and} \  k\in[0,...,n/2-1]\nonumber \\
\nonumber \\
\mbox{and}\ \ \ && Q_{a;k}= Q_{a;k+n/2}= Q_{a +n/2;k}= Q_{a+n;k} = Q_{a;k+n},\label{qa} 
\eeqa
with the conventions : 
\beqa
\alpha_{a+n/2}=\alpha_a,\ \ \ \  \beta_{a+n/2}=\beta_a\ \ \ \mbox{and} \ \ \ m_{a+n/2}=m_a.\label{def}
\eeqa
The structure of  the $P_a$ algebra given by eqs. (\ref{pa}) characterizes completly the algebra of  $Q_{a;k}$. This algebra for equal components $a$ is the only relevant one in the forthcoming analysis. It gives for $a\in[0,...,n/2-2]$ :
\begin{itemize} 
\item For $k\in[0,...,n/2-1]$ :
\beqa
Q^2_{a;k}=-P^2_{a+k} - \frac{m^2_a}{m^2_{\frac{n}{2}-1}}P^2_{\frac{n}{2}-1+k}.\label{qa2}
\eeqa
\item For $k\in[1,...,n/2-1]$ :
\beqa
Q_{a;k}Q_{a;0}&=&\frac{m_a}{m_{\frac{n}{2}-1}}\big[\delta_{a+k,n/2-1}P^2_{\frac{n}{2}-1}
 +  \delta_{a,k-1}P^2_{\frac{n}{2}-1+k} \big] \ \ \mbox{for}\  a\in[0,...,n/2-k-1],\nonumber\\
Q_{a;k}Q_{a;0}&=&\frac{m_a}{m_{\frac{n}{2}-1}}\big[\delta_{a+k,n-1}P^2_{\frac{n}{2}-1}
 +  \delta_{a,k-1}P^2_{\frac{n}{2}-1+k} \big] \ \ \mbox{for}\  a\in[n/2-k,...,n/2-2].\nonumber
\eeqa
\item For $(k>l)\in[1,...,n/2-1]$ :
\beqa
Q_{a;k}Q_{a;l}&=&\frac{m_a}{m_{\frac{n}{2}-1}}\big[\delta_{a+k,l-1}P^2_{\frac{n}{2}-1+l}
\  +\  (k \leftrightarrow l)  \big] \ \ \ \mbox{for}\ \ a\in[0,...,n/2-k-1],\nonumber\\
Q_{a;k}Q_{a;l}&=&\frac{m_a}{m_{\frac{n}{2}-1}}\big[\delta_{a+k,n/2+l-1}P^2_{\frac{n}{2}-1+l}\ + \  (k \leftrightarrow l, l \rightarrow l+n/2) \big]\label{qakl} \\
 &&\ \ \ \ \ \ \ \ \ \ \ \ \mbox{for}\ \ a\in[n/2-k,..., n/2-l-1],\nonumber\\
Q_{a;k}Q_{a;l}&=&\frac{m_a}{m_{\frac{n}{2}-1}}\big[\delta_{a+k,n/2+l-1}P^2_{\frac{n}{2}-1+l}
 \ + \ (k \leftrightarrow l)  \big] \nonumber\\
 &&\ \ \ \ \ \ \ \ \ \ \ \ \mbox{for} \ \  a\in[n/2-l,...,n/2-2].\nonumber
\eeqa
\end{itemize}
In the sequel, eqs. (\ref{mvertex}), (\ref{pa}), (\ref{d2}), and (\ref{qa}) are used to define the $n$(even)-multicomplex extension of any standard vertex operator. Relations (\ref{qakl}) explicitely show the dependence on the subspace dimension $m_a$. For simplicity, we shall work without specifying a representation except when studying particular cases.

\vspace{7mm}
\centerline{\bf Odd multicomplex extension}
\vspace{3mm}

From the above results in the $n$ even case, one can deduce similar expressions for odd $n$ of the multicomplex space. In fact,  the unimodular constraint for odd $n$ is given by :
\vspace{-0.2cm} 
\beqa
\ \ \ \ \ \ \sum_{a=0}^{\frac{n-1}{2}-1}2m_a\beta_a\varphi_a + m_{\frac{n-1}{2}}\beta_{\frac{n-1}{2}}\varphi_{\frac{n-1}{2}} = 0,
\eeqa
and the compact component  $\phi_{\frac{n-1}{2}}$ and generator $P_{\frac{n-1}{2}}$ disappear in all expressions \cite{multisine}. Hence, the following substitutions in the final results available for the even $n$ case may be used to obtain expression valid for odd $n$ :
\beqa
&& (i)\ \   n \rightarrow n+1,\nonumber\\
&& (ii)\ \   m_a \rightarrow 2m_a\ \ \   \mbox{for}\ \ \  a\in[0,..., \frac{n-1}{2}-1],\label{subst} \\
&& (iii)\ \   \alpha_{\frac{n-1}{2}}=0.\nonumber
\eeqa

\vspace{7mm}
\centerline{\bf Operator Product Expansion}
\vspace{3mm}

The calculation of OPE between two different $n$-MC-vertex operators is in keeping with the standard complex case. For later convenience, the following Euclidian propagators related to the kinetic terms are chosen : \ $<\phi_a(z)\phi_b(w)>=-\ln(z-w)\delta_{a,b}$\ \  (for $(a,b)\in[0,...,E(\frac{n}{2})-1]$),\ \ $<\varphi_a(z)\varphi_b(w)>=-\ln(z-w)\delta_{a,b}$\ \ (for $(a,b)\in[0,...,E(\frac{n+1}{2})-2]$) \  and similarly for ${\overline \phi}_a$ and ${\overline \varphi}_a$. It gives (holomorphic part) :
\beqa
J_{{\eta'}^{(k)}}(z) J_{\eta^{(l)}}(w)  \sim (z-w)^{-C^{k,l}(\eta', \eta)} J_{{\eta'}^{(k)}+\eta^{(l)}}(w) + ...\label{ope0}
\eeqa
and similarly for the anti-holomorphic part. For even $n$ :
\beqa
C^{k,l}(\eta', \eta)= {\eta'}^{(k)}.\eta^{(l)}=\sum_{a=0}^{\frac{n}{2}-1}\alpha'_a \alpha_a P_{a+k}P_{a+l} + \sum_{a=0}^{\frac{n}{2}-2}{\beta'}_a \beta_a Q_{a;k} Q_{a;l}\label{scal}
\eeqa
defines the multicomplex extension of the standard scalar product between two (complex components) vectors, with :
\beqa
{\eta}^{(l)}&=& [\alpha_0 P_{l},...,\alpha_{\frac{n}{2}-1} P_{\frac{n}{2}-1+l}; \beta_0 Q_{0;l},...,\beta_{\frac{n}{2}-2} Q_{\frac{n}{2}-2;l}]\label{etak} \\
\mbox{and}\ \ {\eta'}^{(k)}&=& [\alpha'_0 P_{k},...,\alpha'_{\frac{n}{2}-1} P_{\frac{n}{2}-1+k}; \beta'_0 Q_{0;k},...,\beta'_{\frac{n}{2}-2} Q_{\frac{n}{2}-2;k}],\nonumber
\eeqa
two $(n-1)$ (multicomplex components) vectors. As before, $(\{\alpha_a\},\{\beta_a\})$ and $(\{\alpha'_a\},\{\beta'_a\})$  are arbitrary complex parameters. 

From the different symmetries of $(P_{a+k},Q_{a;k})$, $C^{k,l}(\eta', \eta)= C^{l,k}(\eta', \eta)$ and   $C^{k+n,l}(\eta', \eta)= C^{k,l+n}(\eta', \eta)=C^{k+n/2,l+n/2}(\eta', \eta)$. Furthermore, using the structure of the $(P_{a+k},Q_{a;k})$ algebra (\ref{pa}), (\ref{d2}), (\ref{qa}), (\ref{def}), (\ref{qa2}) and (\ref{qakl}), the following result holds for $(k,l)\in[0,...,n/2-1]$:
\beqa
C^{k,k}(\eta', \eta)&=& \sum_{a=0}^{\frac{n}{2}-2}\big[\big(-\alpha'_a \alpha_a +\beta'_a \beta_a \big) \big(-P^2_{a+k}\big)\big]\nonumber\\
&&\ \ \ \ \ \ \ \ \ \ \ \ \ \ \  + \big(-\alpha'_ {\frac{n}{2}-1}\alpha_{\frac{n}{2}-1} + \sum_{a=0}^{\frac{n}{2}-2}\beta'_a \beta_a \frac{m_a^2}{m_{\frac{n}{2}-1}^2}\big) \big(-P^2_{\frac{n}{2}-1+k}\big),\nonumber \\
C^{k,k+n/2}(\eta', \eta)&=& \sum_{a=0}^{\frac{n}{2}-2}\big[\big(\alpha'_a \alpha_a +\beta'_a \beta_a \big) \big(-P^2_{a+k}\big)\big]  \label{Ckl} \\
&&\ \ \ \ \ \ \ \ \ \ \ \ \ \ \   +  \ \big(\alpha'_ {\frac{n}{2}-1}\alpha_{\frac{n}{2}-1} + \sum_{a=0}^{\frac{n}{2}-2}\beta'_a \beta_a \frac{m_a^2}{m_{\frac{n}{2}-1}^2}\big) \big(-P^2_{\frac{n}{2}-1+k}\big),\nonumber\\
C^{k,l}(\eta', \eta)&=& \Big( \sum_{a=0}^{\frac{n}{2}-2}\big[-\beta'_a \beta_a \frac{m_a}{m_{\frac{n}{2}-1}}\delta_{a+k,l-1\ mod(\frac{n}{2})}\big]\Big)\big(-P^2_{\frac{n}{2}-1+l}\big)\nonumber\\
&&\ \ \ \ \ \ \ \ \ \ \ \ \ \ \ \  \ \ \ \ \ \ \ \ \ \ \ \ \ \ \ \ \ \ \ \ \ \ \ \ \ \ \ \ \ \ \ \ \ \ \ \ +     (k \leftrightarrow l),\nonumber\\
C^{k,l+n/2}(\eta', \eta)&=& C^{k,l}(\eta', \eta)\ \ \ \ \ \ \ \ \mbox{for other cases.}\nonumber
\eeqa
$P_{a+n/2}$ being the complex conjugate of $P_a$ (see eqs. (\ref{pa})), it explains the natural change of sign of  $\alpha'_a \alpha_a$ between the two first expressions in eqs. (\ref{Ckl}). If $n$ is odd, the OPE keeps the same form if the substitutions (\ref{subst}) are used.

However, for later convenience, a normalization for MC-vertex operators in keeping with affine Toda field theories will sometimes be used in next sections. Namely, roots $r_i$ for $i=0,...,r$ of affine Lie algebras ($r$ denotes the rank of the algebra) are conventionnally normalized as $r_i^2=2$ for simply laced cases whereas for non-simply laced cases only long roots have this property. As the ATFT potential is expressed in terms of vertex operators \ $\exp\beta r_i.\Phi$ : the associated non-local conserved currents are generally defined by \ $\exp r_i^{\vee}/\beta.\Phi$ with $r_i^{\vee}=2r_i/r_i^2$ denoting the coroots. Moreover, using the linear independence and inversibility of generators $P_a$, eqs. (\ref{Ckl}) indicates that $Det\big(C^{k,k}(\eta', \eta')\big)\neq 0$ only if :
\beqa
&&(i)\ \ \ \  -{\alpha'_a}^2 +{\beta'_a}^2 \neq 0 \ \ \ \ \mbox{for all} \ \ a\in[0,...,E(n/2)-1], \label{invers} \\
&&(ii)\ \ \ \ -{\alpha'}^2_ {\frac{n}{2}-1} + \sum_{a=0}^{\frac{n}{2}-2}{\beta'_a}^2 \frac{m_a^2}{m_{\frac{n}{2}-1}^2} \neq 0\ \ \ \ \ \ \mbox{for even $n$} \nonumber
\eeqa
whereas for $n$ odd, it is sufficient to have eq. $(i)$ in (\ref{invers}) for all \ $a\in [0,...,(n-1)/2-1]$ and $\{\ a\ \ / \ \ \beta'_a\neq 0\}$.

Consequently, $C^{k,k}(\eta', \eta')$ is invertible for even $n$ if parameters $(\{\alpha'_a\},\{\beta'_a\})$ satitisfy (\ref{invers}). Then, in analogy the normalized or dual ``roots'' for the MC-vertex operator are defined as : 
\beqa
{\eta'}^{(k)\vee}=2{\eta'}^{(k)}({\eta'}^{(k)}.{\eta'}^{(k)})^{-1}\label{etaknorm}.
\eeqa
From eqs. (\ref{scal}), (\ref{Ckl}) and (\ref{invers}), we have ${\eta'}^{(k)\vee}=2{\eta'}^{(k)}(C^{k,k}(\eta', \eta'))^{-1}$. Finally, using $P^{-1}_a \sim -P_a$ from eqs. (\ref{pa}), a straightforward calculation using (\ref{Ckl}) gives the expression ${\eta'}^{(k)\vee}$ as in (\ref{etak}) but with the changes :
\beqa
(a) &&\alpha'_a \longrightarrow \frac{-2\alpha'_a}{({\alpha'}_a^2-{\beta'}_a^2)}\ \ \ \  \  \mbox{for}\ \ a \in [0,...,n/2-2],\label{change}\\
(b) &&\beta'_a \longrightarrow \frac{-2\beta'_a}{({\alpha'}_a^2-{\beta'}_a^2)}\ \ \ \  \  \mbox{for}\ \  a \in [0,...,n/2-2], \nonumber \\
(c) &&\alpha'_{n/2-1} \longrightarrow \frac{-2{\alpha'}_{n/2-1}}{({\alpha'}_{n/2-1}^2-\sum_{a=0}^{n/2-2}\frac{m_a^2}{m_{\frac{n}{2}-1}^2}{\beta'}_a^2)},\nonumber \\
(d) && \frac{m_a}{m_{\frac{n}{2}-1}} \longrightarrow \frac{m_a}{m_{\frac{n}{2}-1}}\frac{({\alpha'}_a^2-{\beta'}_a^2)}{({\alpha'}_{n/2-1}^2-\sum_{a=0}^{n/2-2}\frac{m_a^2}{m_{\frac{n}{2}-1}^2}{\beta'}_a^2)}\ \ \ \ \ \ \mbox{in eq.} \ \ \ (\ref{def})\nonumber \\
&&\ \ \ \ \ \ \ \ \ \ \ \ \ \ \ \ \ \ \ \ \ \ \ \ \ \ \ \ \ \ \ \ \ \mbox{for}\ \ \ \  a\in[0,...,n/2-2].\nonumber
\eeqa

For odd $n$, the corresponding results with the substitutions (\ref{subst}).

We can now write the OPE between a ``normalized'' MC-vertex operator with an other ``unnormalized'' MC-vertex operator : 
\beqa
J_{{\eta'}^{(k)\vee}}(z) J_{\eta^{(l)}}(w)  \sim (z-w)^{-C^{k,l}({\eta'}^{\vee}, \eta)} J_{{\eta'}^{(k)\vee}+\eta^{(l)}}(w) + ...\label{ope}
\eeqa
where the multicomplex weights $C^{k,l}({\eta'}^{\vee}, \eta)$ are reported in appendix A (eqs. (\ref{Cklnorm})).

The multicomplex weights $C^{k,l}({\eta'}^{\vee}, {\eta'}^{\vee})$ are given by eqs. (\ref{Ckl}) for \ $\alpha_a\rightarrow\alpha'_a$ and $\beta_a\rightarrow\beta'_a$ with the substitutions (\ref{change}) $(a)$, $(b)$, $(c)$, $(d)$. 

\section{Non-local conserved charges in the multisine-Gordon model}
Using the framework of MC-vertex operators for the multisine-Gordon model, an explicit construction of a set of non-local conserved charges is achieved, preceded by a short review of the model. A brief review of the method is first presented.

\subsection{Perturbed CFT and non-local charges} 
This section presents the analysis used to construct non-local charges and to find their algebraic properties. Consider a conformal field theory perturbed by a relevant scalar operator. Its action writes :
\beqa
{\cal{A}}= {\cal{A}}_{CFT}+ \frac{\lambda}{2\pi}\int d^2z \Phi_{pert.}(z,{\overline z}). 
\eeqa
The dimensionfull parameter $\lambda$ measures the strength of the perturbation away from the critical point ($\lambda=0$).
In the conformal limit, the chiral fields ${\cal{O}}(z,\overline{z})$ and  ${\overline {\cal{O}}}(z,\overline{z})$ satisfy the equations $\partial_{\overline z}{\cal{O}}(z,\overline{z}) = \partial_{z}{\overline {\cal{O}}}(z,\overline{z})=0$. It is then possible to deduce the equations of motion to first order in perturbation theory for the perturbed chiral fields in the perturbed CFT following Zamolodchikov's approach \cite{20} :
\beqa
\partial_{\overline z} {\cal{O}}(z,\overline{z})= \lambda \oint_{z} \frac{dw}{2i\pi}\Phi_{pert.}(w,{\overline z}){\cal{O}}(z,\overline{z}),\label{O} \\
\partial_{z} \overline{{\cal{O}}}(z,\overline{z})= \lambda \oint_{\overline z} \frac{d{\overline w}}{2i\pi}\Phi_{pert.}(z,{\overline w}){\overline{\cal{O}}}(z,\overline{z}).\nonumber
\eeqa

Furthermore, if $J^a(z,{\overline z})$ and ${\overline J}^a(z,{\overline z})$ are identified with local or non-local conserved currents of the perturbed CFT, they should satisfy the following equations of motion to first order in perturbation theory :
\beqa
\partial_{\overline z}J^a(z,{\overline z}) =  \partial_{z}H^a(z,\overline{z}),\label{Ja} \\
\partial_{z}{\overline J}^a(z,\overline{z}) =  \partial_{\overline z}{\overline H}^a(z,\overline{z}).\nonumber
\eeqa
It is clear that these currents become chiral (resp. anti-chiral) at the critical point\,\footnote{For example, it is well known that the stress-energy tensor looses its chirality property in the vicinity of the critical point.}. These currents are chosen to be local with respect to the perturbing field. However, it is seen that eqs.  (\ref{Ja}) can  only be satisfied if the residue of the OPE between the currents and the perturbing field leads to derivatives terms in the r.h.s. of eqs. (\ref{O}). Thus, if this strong constraint is verified and supposing that the perturbing field can be expanded as $\Phi_{pert.}=\Phi^{(0)}_{pert.}+...+\Phi^{(l)}_{pert.}+...$ with $\Phi^{(l)}_{pert.}(z,\overline{z})= \phi^{(l)}_{pert.}(z) {\overline{\phi^{(l)}_{pert.}(z)}}$, one obtains (holomorphic part) :
\beqa
H^a(z,{\overline z}) = \lambda h^a(z) {\overline {\phi^{(l)}_{pert.}(z)}}\ \ \ \ \ \mbox{with}\ \ \partial_z h^a(z) = Res_{z\sim w}(\phi^{(l)}_{pert.}(w)J^a(z))\label{no}
\eeqa
and similarly for the anti-holomorphic part. Finally, the currents generate conserved charges which are defined by :
\beqa
Q^a = \frac{1}{2i\pi}\Big(\oint_z dz J^a + \oint_{\overline z} d{\overline z} H^a\Big) \ \ \ \mbox{and} \ \ \ {\overline Q}^a = \frac{1}{2i\pi}\Big(\oint_{\overline z} d{\overline z} {\overline J}^a + \oint_z dz {\overline H}^a\Big).
\eeqa
Clearly, this approach holds for local or non-local conserved charges. If our interest is in non-local conserved currents, non-trivial relations may appear between them, encoded in their equal-time braided  relations of the form :
\beqa
J^a_\mu(x,t) J^b_\nu(y,t) = R_{dc}^{ab} J^c_\nu(y,t) J^d_\mu(x,t).
\eeqa
Integrability and associativity properties of the theory require the matrix $R_{dc}^{ab}$ to be a solution of the Yang-Baxter equations. Considering now the action of the charges on the physical Hilbert space, similar relations are obtained between the charges and the fields of the theory. In fact, the action of the currents on a field, say $\psi(y)$, is characterized by the integration contour surrounding the point $y$. Moreover, non-locality can be represented by a string attached to the current from $x\rightarrow -\infty$, then, braiding relations arise from the obstructions in moving these strings \cite{17}.  It leads to a non-trivial algebra for the non-local conserved charges characterized by braided commutator relations of the form (described in figure 1):
\beqa
Q^a Q^b - R_{dc}^{ab} Q^c Q^d =  {\hat{Q}}^a (Q^b).\label{contour}
 \eeqa
\vspace{3mm}

\centerline{\epsfig{file=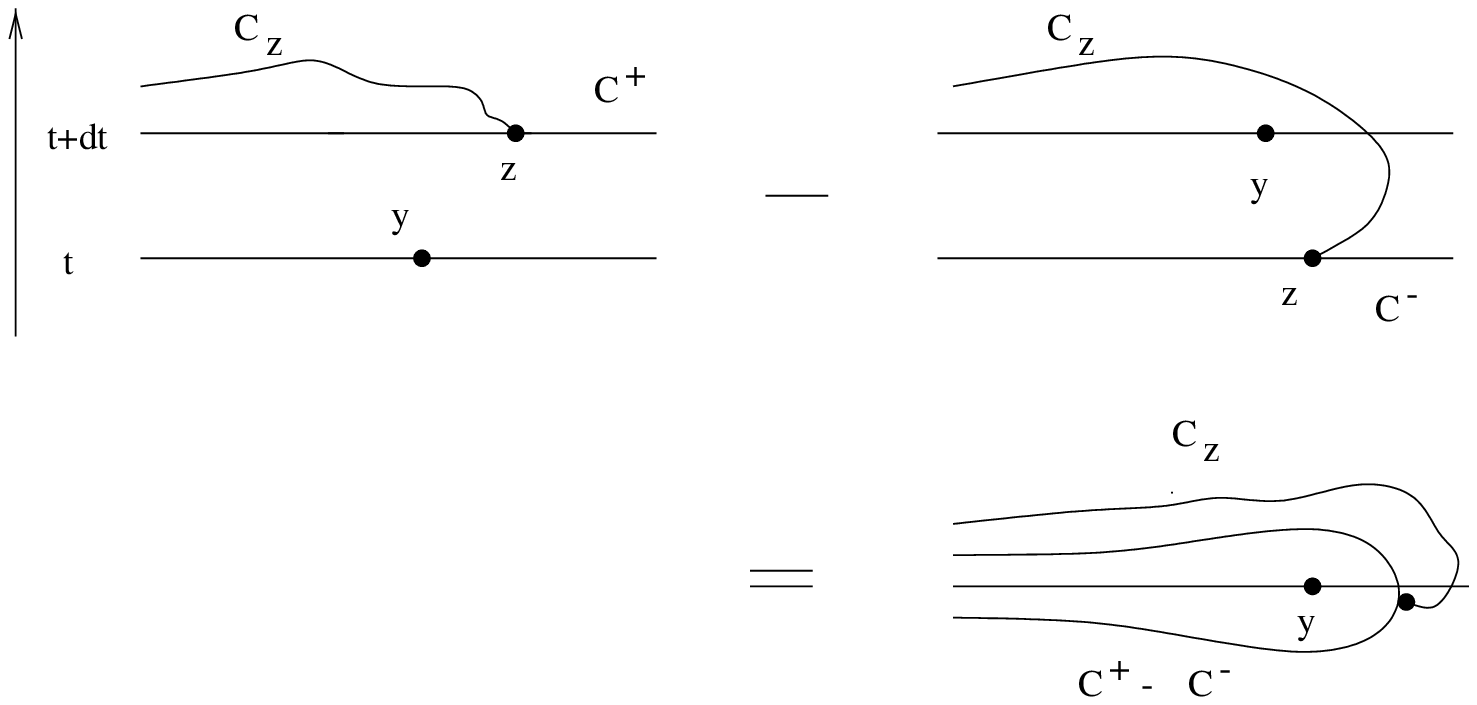,height=70mm,width=120mm}}
\centerline{Figure 1 : equation (\ref{contour})}

\vspace{3mm}
The r.h.s. is always identified as a topological term corresponding to a generalization of the topological extensions of supersymmetry in two dimensions. Finally, the action of the charges on a multiparticle states follows similar rules and is associated to the co-multiplication $\Delta(Q^a)$. We shall identify next and explicitely the different mathematical objects just presented. For more details about these points, see \cite{15,16,17,18}.  

\subsection{Application to the multisine-Gordon model}
As already studied in a previous paper, the multisine-Gordon model based on unimodular multicomplex numbers describes a family of $n-1$ parameters quantum field theories with $n-1$ scalar fields which interact through a multisine potential \cite{multisine,multisine2}. In two dimensions, for some specific values of the coupling parameters\  $(\{\alpha_a\},\{\beta_a\})$\  and of the multicomplex space dimension $n$, the model is integrable at quantum level. However, this model seems not to be integrable classically except for specific values of the parameters\,\footnote{As seen later, it can be identified for $n=4$ to $A_{3}^{(1)}$ ATFT with imaginary coupling constant which possesses classical soliton solutions obtained, for example, by Hirota method \cite{Hirota,Holl}.}.
\vspace{2mm}

{\bf Definition 2} : {\it  The Euclidian action of the multisine-Gordon model associated to the multicomplex algebra\,\footnote{Particulary, as explained in section 2, embeddings $\hbox{\it I\hskip -2.pt M \hskip -7 pt I \hskip - 3.pt \CC}_{n}  \subset \hbox{\it I\hskip -2.pt M \hskip -7 pt I \hskip - 3.pt \CC}_{m}$  may be considered.} $MSG(n|m)$ is defined by :}
\beqa
{\cal{A}}^{(n|m)}&=& \frac{1}{4\pi}\int d^2z \partial_z\Phi\partial_{\overline z}\Phi + \frac{\lambda}{\pi}\int d^2z \mathrm{mus}_0(\Phi),\label{action}\\
 &=& \frac{1}{4\pi}\int d^2z \partial_z\Phi\partial_{\overline z}\Phi + \frac{\lambda}{n\pi}\int d^2z 
\Big(x^{(0)}+...+ x^{(n-1)}\Big)\nonumber  
\eeqa
with :
\beqa
x^{(l)}= \exp\big(\eta^{(l)}.\Phi(z,\overline{z})\big)\label{xl}
\eeqa
where \ $\Phi_j=(\phi_a+{\overline\phi}_a)$\  for\  $(j,a)\in[(0,0),...,(E(n/2-1),E(n/2-1))]$\  and\  $\Phi_j=(\varphi_a+{\overline\varphi}_a)$\  for\  $(j,a)\in[(E(n/2),0),...,(n-2,E\big((n+1)/2\big)-2)]$. For example, in the representation (\ref{rep}) for even $n$ the following expression \cite{multisine,multisine2} is obtained :
\beqa
\label{mus2}
\mathrm{mus}_0(\Phi) = 
{2 \over n}\Big( \sum \limits_{a=0}^{n/2-2}\Big[ 
\cos\alpha_a\phi_a e^{\beta_a\varphi_a}\Big] + \cos\alpha_{n/2-1}\phi_{n/2-1} e^{\big(-\sum_{a=0}^{n/2-2}\beta_a\frac{m_a}{m_{n/2-1}}\varphi_a\big)}\Big).\label{museven}
\eeqa
For odd $n$, a similar expression results with substitutions\footnote{Stricktly speacking, for odd $n$ a coefficient appears before the last vertex operator in (\ref{museven}). However, it can be absorbed by a convenient shift over the fields.} (\ref{subst}).
 
In the deep ultraviolet limit, the mass scale parameter $\lambda$ goes to zero. Normal ordering is used to remove divergences from tadpoles. However, if a fermionic representation\,\footnote{For small $\beta_a$, one can use 2-d fermion-boson correspondence \cite{Col} to rewrite the action (\ref{action}) in a form more suitable for perturbative analysis.} is prefered, depending on the value of $n$, some specific counterterms can be added \cite{multisine,multisine2}, the coupling parameters are thereby unrenormalized. With the above choice of normalization of the kinetic term, the Euclidian propagator is defined as :
\beqa
<\Phi_i(z,{\overline z})\Phi_j(w,{\overline w})>= -2\ln |z-w|\delta_{i,j} \ \ \mbox{for}\ \ (i,j)\in[0,...,n-2]
\eeqa
The action (\ref{action}) is now considered as a perturbation of a conformal field theory in the sense of the previous subsection. In fact, the multisine potential is treated as a perturbation of the conformal field theory associated to $n-1$ free bosons. However, following Zamolodchikov, it is assumed that the space of fields has not been drastically modified by the perturbation. A one-to-one correspondence can then be established between the local operators of the multisine-Gordon model and the fields of the ultra-violet CFT. In the massless limit, each $n-1$ components of the free vectorial boson can be decomposed into one holomorphic and one anti-holomorphic part $\Phi(z,\overline{z})=\phi(z) + \overline{\phi(z)}$. Each MC-vertex operators of the ultraviolet CFT can then be written as a product of two chiral (anti-chiral) MC-vertex operators and of their Virasoro descendents too.

In the spirit above, it follows that the existence of an hypothetical conserved current in the multisine-Gordon model is clearly related to the OPE between this current and the perturbing field. Let us then  use the formalism of MC-vertex operators to construct this quantity.

It is supposed first that the conserved current belongs to the multicomplex space and the expansion of the multisine functions in terms of $x^{(l)}$ is identified as the perturbing field. Using the OPE (\ref{ope0}) for  $J^{(k)}x^{(l)}$ and eqs. (\ref{etak}), (\ref{xl}), the condition  on the residue related to eqs. (\ref{O}), (\ref{Ja}) amounts to a condition on the multicomplex coefficient $C^{k,l}(\eta',\eta)$ given by eqs. (\ref{Ckl}). The task is to find a general set of solutions $(\{\alpha'_a\},\{\beta'_a\})$ - associated to the existence of non-local conserved currents - for the resulting system of constraints. For later convenience (particulary to relate MSG and ATFTs) it is preferable to use ${\eta'}^{(k)\vee}$ (defined in eqs. (\ref{etaknorm}), (\ref{change})) instead of ${\eta'}^{(k)}$ as a possible solution. Thereby, a convenient normalization of ``multicomplex'' roots is ensured, and then of vertex operators as well. In any case, the existence of conserved charges results from the following assumption :

\vspace{2mm}

{\bf Proposition 1} : {\it A set of $2n$ non-local multicomplex conserved currents associated to the multisine-Gordon theories exists if :}
\beqa
&&C^{k,k}(\eta',\eta)=\sum_{a=0}^{n/2-1}2\big(-P_{a}^2\big) = 2,\ \ \mbox{and}\ \ \  C^{k,l}(\eta',\eta)=\sum_{b=0}^{n/2-1} N_b\big(-P_{b}^2\big),\label{propo1} \\
&&\mbox{for $l\neq k$\ \ with}\ \ \ \{N_b\}\in {\mathbb R}_{-}.\nonumber
\eeqa
For even $n$, the form of the constraints is detailed in appendix A where particular solutions, namely ``normalized'' are given, with $(\{\alpha_a\}\in {\mathbb R}^{n/2-1},\{\beta_a\}\in i{\mathbb R}^{n/2-2})$ singled out. Other solutions, called ``unormalized'', can also be obtained as shown in appendix D.\\

\vspace{3mm}
\centerline{\bf Conserved currents for even $n$}
\vspace{2mm}

Choosing an appropriate set of parameters $(\{\alpha'_a\},\{\beta'_a\})$, the only term of the perturbing field which survives corresponds to $l=k(mod (n))$. However, another kind of constraints exists for which the term with $l=k+n/2(mod (n))$ survives, but it corresponds to the change of sign $\alpha'_a \alpha_a \rightarrow -\alpha'_a \alpha_a$ and then gives an isomorphic set of conserved currents, the natural complex conjugate one. After some computations, the equations for the multicomplex non-local currents follow :

\vspace{2mm}

{\bf Proposition 2} : {\it In first order of perturbation theory, the $n$ (holomorphic) non-local multicomplex conserved currents verify :}  
\beqa
\partial_{\overline z}J^{(k)} = \partial_z H^{(k)} &&\ \ \ \ \ \ \mbox{for}\ \ \ k\in[0,...,n-1] \nonumber \\
\nonumber \\ 
  \mbox{with} && \ J^{(k)} = \exp\big({\eta'}^{(k)}.\phi(z)\big),\label{h}\\
  &&  \  H^{(k)}= \frac{2\lambda}{n}\Big(\exp\big(({\eta'}^{(k)}+ \eta^{(k)}).\phi(z)+ \eta^{(k)}.{\overline {\phi(z)}} \big)\Big)\nonumber
\eeqa 
and similarly for the anti-holomorphic part with ($\phi \leftrightarrow \overline{\phi}$). Here ${\eta'}^{(k)}$ is given by eqs. (\ref{etak}) (for ``unormalized'' solutions) or eqs. (\ref{etak}), (\ref{etaknorm}), (\ref{change}) (${\eta'}^{(k)}\rightarrow{\eta'}^{(k)\vee}$ for ``normalized solutions'') and similarly for the anti-holomorphic part; $\eta^{(k)}$ is given by eqs. (\ref{etak}).

\vspace{2mm}

{\bf Definition 3} : {\it $2n$ non-local multicomplex conserved charges generated by $2n$ non-local multicomplex conserved currents are defined by :}
\beqa
Q^{(k)} &=& \frac{1}{2i\pi} \Big( \oint_z dz J^{(k)}\ + \ \oint_{\overline z} d{\overline z} H^{(k)}\Big)\ \ \ \ \mbox{for}\ \ k\in[0,...,n-1],\nonumber \\
{\overline Q}^{(k)} &=& \frac{1}{2i\pi} \Big(\oint_{\overline z} d{\overline z} {\overline J}^{(k)} \ +\ \oint_z dz {\overline H}^{(k)}\Big)\label{mucur}
\eeqa 
with $J^{(k)}$ and $H^{(k)}$ (resp. ${\overline J}^{(k)}$ and ${\overline H}^{(k)}$) given in eqs. (\ref{h}).

The non-locality of the multicomplex conserved charges results from the massless limit where $\partial_{\overline z}\phi_j=\partial_{z}{\overline \phi_j}=0$ and since $\partial_z=\partial_x+\partial_t$ and $\partial_{\overline z}=\partial_x-\partial_t$, it follows that $(\partial_x+\partial_t)\Phi_j=(\partial_x+\partial_t)\phi_j=2\partial_x \phi_j$. This last equation gives the expression of the fields ($\phi_j, {\overline \phi_j}$) in a non-local way \,\footnote{The term $\int_{-\infty}^{x}dy\partial_t\Phi_j(y,t)$ exhibits the string which is attached to the current $J^{(k)}$ through the fundamental field $\Phi_j$} in terms of the MSG fundamental field $\Phi_j$:
\beqa
\phi_j(x,t) &=& \frac{1}{2}\Big( \Phi_j(x,t) + \int_{-\infty}^{x}dy\partial_t\Phi_j(y,t) \Big) \ \ \ \mbox{for}\ \ j\in[0,...,n-2], \nonumber \\
{\overline \phi_j}(x,t) &=& \frac{1}{2}\Big( \Phi_j(x,t) - \int_{-\infty}^{x} dy\partial_t\Phi_j(y,t) \Big).\label{phij}
\eeqa
As explained above, eventhough these non-local expressions are derived in the massless limit, they can be taken to define the chiral components ($\phi_j,{\overline \phi}_j$) in the perturbed theory.

From the expressions of vector (chiral, anti-chiral) fields ($\phi,{\overline \phi}$) in terms of the MSG bosonic vector field $\Phi$ (\ref{phij}), of the equal-time canonical commutation relation 
\beqa
[\phi_i(x,t),{\overline \phi_j}(y,t)] = -i\pi\delta_{i,j}\ \ \ \  \mbox{for all}\ (x,y)
\eeqa
(since \ \ $[\Phi_i(x,t), \partial_t\Phi_j(y,t)]=4i\pi\delta_{i,j}\delta(x-y)$), expressions (\ref{Ckl}) and the abelian property of the multicomplex algebra (\ref{pa}), the braiding properties of non-local multicomplex currents are found as :
\beqa
J^{(k)}(x,t){\overline J}^{(l)}(y,t) = \exp(-i\pi C^{k,l}(\eta',\eta')){\overline J}^{(l)}(y,t) J^{(k)}(x,t)\label{braid}
\eeqa
for all\ \ \ $(x,y)$\ \ and \ \ $(k,l)\in[0,...,n-1]$.

Along similar lines and using constraint equations, it can be shown that :

\vspace{2mm}

{\bf Proposition 3} : {\it Braiding relations between $J^{(k)}, {\overline H}^{(l)}$ and  $H^{(k)}, {\overline J}^{(l)}$ are identical to} (\ref{braid}) {\it iff :}
\beqa
C^{k,l}(\eta',\eta)=\sum_{b=0}^{n/2-1}\Big[N_b\big(-P_{b}^2\big)\Big]\ \ \ \mbox{for $l\neq k$,\ \ with}\ \ \ \{N_b\}\in{\mathbb Z}_{-}.\label{C}
\eeqa
To first order in perturbation theory, the calculation of the multicomplex topological term ${\hat Q}^{(k)}({\overline Q}^{(l)})$ exhibits only diagonal $(l=k)$ contributions. It writes explicitely (compared to eq. (\ref{no}), with $\lambda\rightarrow \frac{2}{n}\lambda$) : 
\beqa
 {\hat Q}^{(k)}({\overline Q}^{(l)})&=&\delta_{k,l}T^{(k)}= \delta_{k,l}\frac{\lambda}{ni\pi} \int_t \big( dz\partial_z + d{\overline z} \partial_{\overline z}\big) h^{(k)}(z) {\overline {h^{(k)}({z})}}\label{topo} \\
\mbox{with}\ \ \ \ && h^{(k)}= \exp\big(({\eta'}^{(k)}+\eta^{(k)}).\phi(z)\big),\nonumber \\
&& {\overline {h^{(k)}}}= \exp\big(({\eta'}^{(k)}+\eta^{(k)}).{\overline {\phi(z)}}\big),\nonumber
\eeqa
e.g. :
\beqa
\delta_{k,l} T^{(k)}= \frac{\lambda}{in\pi} \int_t dx\partial_x \exp\big(({\eta'}^{(k)}+\eta^{(k)}).\Phi\big).\label{Tmulti}
\eeqa
Using (\ref{h}), (\ref{mucur}), (\ref{braid}), (\ref{topo}) and assuming relation (\ref{C}) to be satisfied :

\vspace{2mm}

{\bf Proposition 4} : {\it Iff parameters} $(\{\alpha'_a\},\{\beta'_a\})$ {\it satisfy eqs.} (\ref{propo1}) {\it and} (\ref{C}), {\it the algebra of the $2n$ non-local conserved multicomplex charges is given by :} 
\beqa
&& Q^{(k)}{\overline Q}^{(l)} - q^{-2}_{k,l}{\overline Q}^{(l)} Q^{(k)} = \delta_{k,l}T^{(k)}\label{mualg} \\
\nonumber \\
&&\mbox{with the deformation} \ \ \ \  q^{-2}_{k,l} = \exp(-i\pi C^{k,l}(\eta',\eta')).\nonumber
\eeqa
Clearly the whole analysis is independent of any choice of representation for the multicomplex algebra. In fact, at this point the MSG model appears as a quantum field theory formulated in terms of multicomplex generators. The multicomplex conserved charges have then also generic MC-expressions. Different solutions to the constraint equations are now envisaged.

\vspace{5mm}
\centerline{\bf Restriction of parameters space}

\begin{itemize}
\item {\it Restriction to} $(\{\alpha_a\}\in{\mathbb R}^{*n/2-1},\{\beta_a\}\in i{\mathbb R}^{*n/2-2})$ {\it for even} $n$
\end{itemize}
\vspace{-3mm}
For this choice, several ATFTs enter in our study\,\footnote{In fact, due to the structure of the multicomplex algebra and parameters dependence, all affine Lie algebras (associated to the ATFT potential) can not obtained.}\cite{multisine2}. As explained in appendix A, $(\{\alpha'_a\},\{\beta'_a\})$ should verify constraints related to eqs. (\ref{cou}) - for non-local conserved charges to exist -.  In appendix A, a trivial solution is also given with\ \ ${\eta'}^{(k)}={\eta}^{(k)\vee}=2{\eta}^{(k)}({\eta}^{(k)})^{-2}$,\  e.g. \ $\alpha_a'=\alpha_a;\ \beta_a'=\beta_a$ if (\ref{C}) is satisfied. Then, it follows :

\vspace{2mm}

{\bf Proposition 5} : {\it For $(\{\alpha_a\}\in{\mathbb R}^{*n/2-1},\{\beta_a\}\in i{\mathbb R}^{*n/2-2})$, a ``normalized'' set of non-local multicomplex conserved charges exists if :}
\beqa
&&C_{a,a+n/2}\big(\eta^{\vee},\eta\big)\in {\mathbb Z}_{-} \ \ \ \ \ \mbox{and}\ \ \ \  \ 
C_{n/2-1,n-1}\big(\eta^{\vee},\eta\big)\in {\mathbb Z}_{-},\label{c3}\\
&&C_{a,n/2-1}\big(\eta^{\vee},\eta\big)\in {\mathbb Z}_{-} \ \ \ \ \ \mbox{and}\ \ \ \  \ 
C_{n/2-1,a}\big(\eta^{\vee},\eta\big)\in {\mathbb Z}_{-}.\nonumber
\eeqa
where the coefficients $C_{a,b}\big(\eta^{\vee},\eta\big)$ are given in appendix A.

Some simplifications occurs in the r.h.s. of expression (\ref{mualg}). First, with a field $\Phi$ configuration such that $\Phi(x\rightarrow\infty,t)=0$, one may write :
\beqa
\int_t dx\partial_x \exp\big(({\eta}^{(k)\vee}+\eta^{(k)}).\Phi\big)=\Big(1-\exp\big(-\ ({\eta}^{(k)\vee}+\eta^{(k)}).\int_t dx\partial_x\Phi\big)\Big).
\eeqa

\vspace{2mm}

{\bf Definition 4} : {\it The multicomplex topological charge is defined as :}
\beqa
{\cal{T}}^{(k)}= -\frac{\eta^{(k)}}{2i\pi}.\int_t dx\partial_x\Phi.\label{sub}
\eeqa
Eq. (\ref{Tmulti}) reduces to :
\beqa
\delta_{k,l} T^{(k)}= \frac{\lambda}{in\pi} \Big[ 1-\exp\big(2i\pi(2(C^{k,k})^{-1}(\eta,\eta)+1){\cal{T}}^{(k)}\big)\Big].\label{sub2}
\eeqa
A field normalization can be chosen\,\footnote{For the SG model, the normalization condition is fixed by the periodicity of the perturbing term $\cos(\alpha_0\Phi)$.} such that $e^{(2i\pi{\cal{T}}^{(k)})}\sim 1$ provided there is no problem around the boundaries. Consequently :
\beqa
q_{k} = \exp(2i\pi (C^{k,k})^{-1}(\eta,\eta))= \exp(i\pi/2 (C^{k,k})(\eta^{\vee},\eta^{\vee})),\label{deformation}
\eeqa
eq. (\ref{mualg}) writes : 
\beqa
Q^{(k)}{\overline Q}^{(l)} - q_{k}^{-C^{l,k}(\eta^{\vee},\eta)}{\overline Q}^{(l)} Q^{(k)} = \delta_{k,l}\frac{\lambda}{in\pi} \Big[ 1-q_{k}^{2{\cal{T}}^{(k)}}\Big].\label{Qalg}
\eeqa
where :
\beqa
&&C^{l,k}(\eta^{\vee},\eta)=C^{k,l}(\eta,\eta^{\vee})\\
&&q_{l}^{-C^{k,l}(\eta^{\vee},\eta)}=q_{k}^{-C^{l,k}(\eta^{\vee},\eta)}=q_{k,l}^{-2}\nonumber
\eeqa

The value of the topological charge can then be found for any MC-vertex operator of the form : 
\beqa
V^{(l)}=\exp\big(\xi^{(l)}.\phi + {\overline \xi}^{(l)}.{\overline \phi}\big)
\eeqa
from (\ref{sub}). It is easy to show that :
\beqa
{\cal{T}}^{(k)}V^{(l)}&=&\big[\eta^{(k)}.(\xi^{(l)} - {\overline \xi}^{(l)})\big]V^{(l)}\\
&=&\big[C^{k,l}(\eta,\xi)-C^{k,l}(\eta,{\overline \xi})\big]V^{(l)}.\nonumber
\eeqa
In particular, for $V^{(l)}\in\big(J^{(l)},H^{(l)}\big)$ and with (\ref{h}), the topological charge of the non-local multicomplex conserved charges writes now (an opposite sign occurs for the anti-holomorphic part) : 
\beqa
\big[{\cal{T}}^{(k)},Q^{(l)}\big]&=&\ \ \ C^{l,k}(\eta^{\vee},\eta)\ Q^{(l)},\label{valtopo}\\
\big[{\cal{T}}^{(k)},{\overline Q}^{(l)}\big]&=&-\ C^{l,k}(\eta^{\vee},\eta)\ {\overline Q}^{(l)}.\nonumber
\eeqa
Hence, it follows that :

\vspace{2mm}

{\bf Proposition 6} : {\it For $\{\alpha_a\}\in \mathbb{R}^{*n/2-1}$ and $\{\beta_a\}\in i\mathbb{R}^{*n/2-2}$, eqs.} (\ref{deformation}), (\ref{Qalg}) {\it and} (\ref{valtopo}) {\it with assumption} (\ref{C}) {\it describe the algebra of non-local multicomplex charges of the MSG model to lowest order in perturbation theory iff eqs.} (\ref{propo1}) {\it and} (\ref{C}) {\it are satisfied.}

\vspace{4mm}

Using a scaling argument, it can be checked that this algebra is exact to all order in perturbation theory (following \cite{16}).

In a forthcoming publication we shall discuss how this algebra can be obtained as one basis of the natural multicomplex extension of $U_q({\hat {sl_2}}(\CC))$.

\subsection{Representation of MSG theories and non-local charges algebra (even $n$)}
In \cite{multisine,multisine2}, our attention focused on the MSG model in the representation (\ref{rep}) and its numerous links with known QFTs. A few remarks are in order to keep only relevant informations from multicomplex non-local charges construction given by eqs. (\ref{mucur}) if we consider the model associated to this representation. The algebra of non-local conserved charges is obtained as follows : representation (\ref{rep}) is retained (the action (\ref{action}) is real-valued since only the complex number $i$ appears through $\cos\alpha_a\phi_a$) and the aim is to obtain directly the charges in terms of standard complex numbers. In other words, for the representation (\ref{rep}), the multisine function is clearly multiplied by the $(n\times n)$ identity matrix. Then, in this representation, each component of the non-local multicomplex charges verifies the algebra (\ref{mualg}). It is easy to see, from eqs. (\ref{Ckl}), (\ref{mucur})  that the component $(00)$ of the multicomplex charge $Q^{(k-j)}$ verifies the same algebra as that of the component $(jj)$ of the multicomplex charge $Q^{(k)}$ for $(k,j)\in[0,...,n/2-1]$ (similarly for the anti-holomorphic charges). Therefore, an isomorphism between these different components exists which allows keeping all the same one component of all the non-local multicomplex charges as a set of non-local conserved charges of the MSG theory in this representation. Hence, the choice of a specific component is irrelevant here.

A complete set of non-local conserved charges of the MSG in representation (\ref{rep}) can then be built since the restriction of the space of non-local multicomplex charges reduces to $n$ copies of it.

For even $n$, let us still consider the representation $(\ref{rep})$ of the multicomplex algebra. The (diagonal) components verify : 
\beqa
(\pi\big[P_{a}^2\big])_{jj}&=&-\delta_{a,j} \ \ \ \mbox{for}\ \ \ (a,j)\in[0,...,n/2-1].   
\eeqa
Only one component, say the first, $(jj)=(00)$ can be retained. Defining the non-local conserved charges of the MSG as ${\cal Q}^{(a)}$ for $\big(\pi\big[Q^{(n/2-a)}\big]\big)_{00}$ associated to ${\cal J}^{(a)}$ for $\big(\pi\big[J^{(n/2-a)}\big]\big)_{00}$and the topological term as $T^{(a)}$ for $\big(\pi\big[T^{(n/2-a)}\big]\big)_{00}$, the algebra for $(a,b)\in[0,...,n-1]$ is such that :
\beqa
{\cal Q}^{(a)}{\overline {\cal Q}}^{(b)} - q^{-C_{a,b}(\eta,\eta^{\vee})}_{a}{\overline {\cal Q}}^{(b)} {\cal Q}^{(a)} &=& \delta_{a,b}T^{(a)} \nonumber
\eeqa
with $q_a=\exp(i\pi/2C_{a,a}(\eta^{\vee},\eta^{\vee}))$, $C_{a,b}(\eta,\eta^{\vee})$ reported in appendix C (eqs. (\ref{Cab})). Here ${\cal Q}^{(a)}$, ${\overline {\cal Q}}^{(a)}$ are defined as follows, depending on the basis of non-local conserved currents (``normalized'' or ``unnormalized'').\\

\vspace{3mm}
\centerline{\bf Deformed algebra associated to ``normalized'' solutions} 
\vspace{1mm}
\centerline{\bf for ($\{\alpha_a\}\in \mathbb{R}^{*n/2-1}$, $\{\beta_a\}\in i\mathbb{R}^{*n/2-2}$)}
\vspace{3mm}
As shown in appendix A, the non-local {\it multicomplex} currents $J_{\eta^{(k)\vee}}$ are conserved for all $k$ with eq. (\ref{c3}) satisfied. Then, in the representation (\ref{rep}),  ${\cal Q}^{(a)}$ is obtained for :
\beqa
{\cal J}^{(a)}&=&\exp\big(i\frac{2{\alpha_a}\phi_a}{({\alpha^2_a}-{\beta^2_a})} - \frac{2{\beta_a}\varphi_a}{({\alpha^2_a}-{\beta^2_a})}\big)\ \ \ \ \ \mbox{for}\ \ a\in[0,...,n/2-2],\\
{\cal J}^{(n/2-1)}&=&\exp\big(i\frac{2{\alpha}_{n/2-1}\phi_{n/2-1}}{({\alpha}_{n/2-1}^2-\sum_{a=0}^{n/2-2}\frac{m_a^2}{m_{\frac{n}{2}-1}^2}{\beta}_a^2)} + \frac{(\sum_{a=0}^{n/2-2}2{\beta}_a \frac{m_a}{m_{\frac{n}{2}-1}}\varphi_a)}{({\alpha}_{n/2-1}^2-\sum_{a=0}^{n/2-2}\frac{m_a^2}{m_{\frac{n}{2}-1}^2}{\beta}_a^2)}\big)\nonumber
\eeqa
with the deformations \ \ \ \  $(q^2_{a,b}=q^2_{b,a})$ : 
\beqa
&&q^2_{a,a}=q^2_{a+n/2,a+n/2}=\exp\big(-i\pi\frac{4}{({\alpha}_a^2-{\beta}_a^2)} \big),\ \ \ \ \ \ \ \ \ \ q^2_{a,a+n/2}= \exp\big(i\pi\frac{4( {\alpha}^2_{a} + {\beta}^2_{a} )}{({\alpha}_a^2-{\beta}_a^2)^2}\big),\label{alg} \nonumber \\
&&q^2_{n/2-1,n/2-1}= \exp\big(-i\pi\frac{4}{({\alpha}_{n/2-1}^2-\sum_{a=0}^{n/2-2}\frac{m_a^2}{m_{\frac{n}{2}-1}^2}{\beta}_a^2)} \big), \nonumber 
\eeqa
\beqa
&&q^2_{n/2-1,n-1}= \exp\Big(i\pi\frac{4({\alpha}_{n/2-1}^2+\sum_{a=0}^{n/2-2}\frac{m_a^2}{m_{\frac{n}{2}-1}^2}{\beta}_a^2)}{({\alpha}_{n/2-1}^2-\sum_{a=0}^{n/2-2}\frac{m_a^2}{m_{\frac{n}{2}-1}^2}{\beta}_a^2)^2} \Big), \\
&&q^2_{n/2-1,a}=q^2_{n-1,a}= \exp\big(-i\pi\frac{2{\beta}^2_{a}}{({\alpha}_a^2-{\beta}_a^2)}\frac{2}{({\alpha}_{n/2-1}^2-\sum_{a=0}^{n/2-2}\frac{m_a^2}{m_{\frac{n}{2}-1}^2}{\beta}_a^2)} \frac{m_{a}}{m_{n/2-1}}\big), \nonumber \\
&&q^2_{n/2-1,a+n/2}=q^2_{n-1,a+n/2}, \nonumber \\
&&q^2_{a,b}=1 \ \ ,\ \ \ \ \mbox{in other cases, with}\ \ \  b\in[0,...,n/2-1].\nonumber
\eeqa
Along similar lines and using eqs. (\ref{Tmulti}), the topological term for $a\in[0,...,n/2-2]$ is given by :
\beqa
T^{(a)}&=& \frac{\lambda}{in\pi} \int_t dx\partial_x \exp\Big( -i\big(\frac{-2{\alpha}_{a}}{({\alpha}_a^2-{\beta}_a^2)}+ {\alpha}_{a}\big).(\phi +{\overline \phi})_{a}\label{topo2} \\
&&\ \ \ \ \ \ \ \ \ \ \ \ \ \ \ \ \ \ +  \big(\frac{-2{\beta}_{a}}{({\alpha}_a^2-{\beta}_a^2)}+ {\beta}_{a}\big).(\varphi +{\overline \varphi})_{a}\Big)\nonumber
\eeqa
and
\beqa
T^{(n/2-1)}&=&\frac{\lambda}{in\pi} \int_t dx\partial_x \exp\Big( -i\big(\frac{-2{\alpha}_{n/2-1}}{({\alpha}_{n/2-1}^2-\sum_{a=0}^{n/2-2}\frac{m_a^2}{m_{\frac{n}{2}-1}^2}{\beta}_a^2)}+ {\alpha}_{n/2-1}\big).(\phi +{\overline \phi})_{n/2-1}\label{topon/2-1}\nonumber \\
&& - \ \sum_{a=0}^{n/2-2}\Big(\big(\frac{-2{\beta}_a}{({\alpha}_{n/2-1}^2-\sum_{a=0}^{n/2-2}\frac{m_a^2}{m_{\frac{n}{2}-1}^2}{\beta}_a^2)}+\beta_a\big)\frac{m_a}{m_{n/2-1}}\Big).(\varphi +{\overline \varphi})_{a}\Big).
\eeqa
From the parameters space restriction, the topological charge is easily obtained in the same way from (\ref{sub}) :
\beqa
{\cal T}^{(a)}&=& \frac{1}{2i\pi} \int_t dx\partial_x \Big( i{\alpha}_{a}.(\phi +{\overline \phi})_{a}\label{topo4} - {\beta}_{a}.(\varphi +{\overline \varphi})_{a}\Big)\nonumber
\eeqa
and 
\beqa
{\cal T}^{(n/2-1)}=\frac{1}{2i\pi} \int_t dx\partial_x \Big( i{\alpha}_{n/2-1}.(\phi +{\overline \phi})_{n/2-1}\label{topo5} +  \sum_{a=0}^{n/2-2}\frac{m_a}{m_{n/2-1}}{\beta}_{a}.(\varphi +{\overline \varphi})_{a}\Big).
\eeqa
Finally, it is worth noting that :  
\beqa
{\cal Q}^{(a+n/2)}=({\cal Q}^{(a)})_{|(\alpha_a)\rightarrow(-\alpha_a)},\\
T^{(a+n/2)}=(T^{(a)})_{|(\alpha_a)\rightarrow(-\alpha_a)}.
\nonumber
\eeqa

All these results can be extended to the family of QFTs for odd $n$ values : this is reported in appendix C since it comes directly from straightforward restrictions of the even $n$ case. In the next section several applications are given, showing perfect agreement with known results, ATFTs ones for example.

\section{Examples}
Different examples are now treated using representation (\ref{rep}). For $n=2$, the model is denoted $MSG_{(2|2)}(\alpha_0)$ : there is only one generator \ $P_0$ \ as \ $P_0^2=-1$\  and it is isomorphic to the standard complex number. From eqs. (\ref{etak}), (\ref{action}) and (\ref{xl}), this model corresponds to the standard sine-Gordon model, with only one parameter $\alpha_0$, all the non-compact parameters $\beta_a$ disappearing. In this case, the system of constraints related to eqs. (\ref{cou}),  (\ref{op1}) and (\ref{Cklnorm}) gives the trivial solution $\alpha'_0=\alpha_0$. Using (\ref{alg}), the following algebra for the non-local charges is immediate :
\beqa
{\cal Q}^{(0)}{\overline {\cal Q}}^{(0)}- q^{-2}{\overline {\cal Q}}^{(0)} {\cal Q}^{(0)}&=& T^{(0)}, \nonumber \\
{\cal Q}^{(1)}{\overline {\cal Q}}^{(1)} - q^{-2}{\overline {\cal Q}}^{(1)} {\cal Q}^{(1)} &=& T^{(1)},\label{algsine} \\
{\cal Q}^{(0)}{\overline {\cal Q}}^{(1)} - q^{2}{\overline {\cal Q}}^{(1)} {\cal Q}^{(0)} &=& 0, \nonumber \\
{\cal Q}^{(1)}{\overline {\cal Q}}^{(0)} - q^{2}{\overline {\cal Q}}^{(0)} {\cal Q}^{(1)} &=& 0, \nonumber \\
\nonumber \\
 \mbox{with}\ \ \ \ q=\exp(-2i\pi/\alpha_0^2).\label{q}&&
\eeqa
As noticed above,\  $T^{(1)}$\  is the standard complex conjugate of\  $T^{(0)}$ which writes, from eq. (\ref{topo2}) :
\beqa
T^{(0)}= \frac{\lambda}{2i\pi} \int_t dx\partial_x \exp\big( i(\frac{2}{\alpha_0}- \alpha_0)(\phi_0+{\overline\phi_0})\big),
\eeqa
in full agreement with the results of D. Bernard and A. LeClair \cite{16} concerning the sine-Gordon model, with the identification : ${\cal Q}^{(0)}=  {\cal Q}^{+}\ \ ; \  \  {\overline {\cal Q}}^{(0)}  = {\overline {\cal Q}}^{-} \  \  ; \  \  {\cal Q}^{(1)}=  {\cal Q}^{-} \  \  ; \  \  {\overline {\cal Q}}^{(1)} = {\overline {\cal Q}}^{+}$. This algebra is identified as the infinite dimensional quantum envelopping algebra $U_q({\hat {sl_2}})$.  

\subsection{$MSG_{(3|m)}(\alpha_0;\beta_0)$}
For specific parameters and $m=3,4$, these models are integrable \cite{multisine,multisine2}. The non-local charges obtained for these QFTs can be seen as a restriction to $\alpha_1=0$ of the general $MSG_{(4|m)}(\alpha_0,\alpha_1;\beta_0)$ model studied in the next subsection. For $n=3$ their associated algebra is described by eqs. (\ref{algodd}) of appendix B. It can be easily checked that different parameters lead to known ATFTs. For instance, the non-local charges algebras of models $MSG_{(3|3)}(\beta\frac{\sqrt 3}{\sqrt 2};i\beta\frac{1}{\sqrt 2})$, $MSG_{(3|3)}(\beta;i\beta)$ and $MSG_{(3|4)}(\beta{\sqrt 2};i\beta{\sqrt 2})$ are respectively identified to $U_q(A_2^{(1)})$, $U_q(C_2^{(1)})$ and $U_q(B_2^{(2)})$. These results are in perfect agreement with those expected for ATFTs. Explicitely, the deformations simplifies to : 
\beqa
q^{-2}_{a,b}=\big(q^{-\frac{1}{2}r_a^2}\big)^{C_{a,b}}\ \ \ \mbox{with}\ \ q=\exp(-\frac{i\pi}{\beta^2}),\label{qab}
\eeqa
where $r_a\ (a=0,1,2)$ denote simple roots\,\footnote{$r_0$ is not necessarely identified to the affine extension} and $C_{a,b}=C_{a,b}(\eta,\eta^{\vee})=C_{b,a}(\eta^{\vee},\eta)$ (see eqs. (\ref{Cab})) different elements of the extended Cartan matrix associated respectively to $A_2^{(1)}$, $C_2^{(1)}$ and $B_2^{(2)}$ affine Lie algebras. Some remarks about the expression of the topological charge in terms of Cartan subalgebras are common to those in the next subsection for $n=4$. 

\subsection{$MSG_{(4|m)}(\alpha_0,\alpha_1;\beta_0)$}
The QFT corresponding to the case $n=4$ has three parameters $\alpha_0,\alpha_1, \beta_0$. For normalized MC-vertex operators,  the resulting deformations associated to the algebra of the $8$ non-local conserved charges are (see (\ref{alg})): 
\beqa
q^2_{00}=q^2_{22}=\exp\big(-i\pi\frac{4}{({\alpha}_0^2-{\beta}_0^2)}\big),\ \ \ \ q^2_{02}=\exp\big(i\pi\frac{4({\alpha}_0^2+{\beta}_0^2)}{({\alpha}_0^2-{\beta}_0^2)^2}\big),\nonumber \\
q^2_{11}=q^2_{33}=\exp\big(-i\pi\frac{4}{({\alpha}_1^2-{\beta}_0^2\frac{m_0^2}{m_{1}^2})}\big),\ \ \ \ \ q^2_{13}=\exp\big(i\pi\frac{4({\alpha}_1^2+{\beta}_0^2\frac{m_0^2}{m_{1}^2})}{({\alpha}_1^2-{\beta}_0^2\frac{m_0^2}{m_{1}^2})^2}\big),\nonumber \\
q^2_{10}=q^2_{30}= q^2_{12}=q^2_{32}= \exp\big(-i\pi\frac{4{\beta}_0^2}{({\alpha}_0^2-{\beta}_0^2)({\alpha}_1^2-{\beta}_0^2\frac{m_0^2}{m_{1}^2})}\frac{m_0}{m_{1}})\big).\nonumber
\eeqa
and the topological terms are given by :
\beqa
T^{(0)}&=&\frac{\lambda}{4i\pi}\int_t dx\partial_x \Big( \exp\big(-i(\alpha_0+\frac{-2\alpha_0}{({\alpha}_0^2-{\beta}_0^2)})(\phi_0+{\overline\phi_0})\nonumber\\
&&\ \ \ \ \ \ \ \ \ \ \ \ \ \ \ \ \ \ \ \ \ \ + \ (\beta_0+\frac{-2\beta_0}{({\alpha}_0^2-{\beta}_0^2)})(\varphi_0+{\overline\varphi_0}) \big)\Big)\nonumber
\eeqa
and
\beqa
T^{(1)}&=&\frac{\lambda}{4i\pi}\int_t dx\partial_x \Big( \exp\big(-i(\alpha_1+\frac{-2\alpha_1}{({\alpha}_1^2-{\beta}_0^2\frac{m_0^2}{m_1^2})})(\phi_1+{\overline\phi_1})\nonumber \\
&&\ \ \ \ \ \ \ \ \ \ \ \ \ \ \ \ \ \ \ \ \ \ - \ (\beta_0+\frac{-2\beta_0}{({\alpha}_1^2-{\beta}_0^2\frac{m_0^2}{m_1^2})})\frac{m_0}{m_1}(\varphi_0+{\overline\varphi_0}) \big)\Big).\nonumber
\eeqa
 
\subsubsection{$MSG_{(4|4)}(\alpha_0,\alpha_1;\beta_0)$}
From eqs. (\ref{action}) and (\ref{museven}), the QFT writes :
\begin{eqnarray}
{\cal{A}}^{(4|4)}&=& \int d^2x \frac{1}{8\pi} \big( (\partial_{\mu} \varphi_0)^2 + (\partial_{\mu} \phi_0)^2 + (\partial_{\mu} 
\phi_1)^2\big) \\ \nonumber
&& \ \  \ \ \ \ -\frac{\lambda}{4\pi} \left( \exp(\beta_0\varphi_0)
\cos(\alpha_0\phi_0) + \exp(-\beta_0\varphi_0)\cos(\alpha_1\phi_1) 
\right)\Big] .\label{QFT3}
\end{eqnarray}
Its connexion with the multicomplex formalism is discussed in \cite{multisine}. It has a $U(1)\otimes U(1)$ symmetry and its dynamics for $\alpha_0=\alpha_1$ was studied in \cite{Fateev1}. Moreover, this model possesses a dual representation in terms of an integrable deformation of the non-linear $\sigma$ model.

Let us now study particular cases and their underlying hidden symmetries associated to the present approach. 
\begin{itemize}
\item $MSG_{(4|4)}(\alpha_0,\alpha_1;0)$\\
For $\beta_0=0$, this model reduces to two decoupled SG. For ``normalized'' currents, constraint equations gives : \ $\alpha_0' =\alpha_0$\ and $\ \alpha_1'=\alpha_1$. The algebra of non-local conserved charges obtained in (\ref{alg}) simply decouples as a product of two subalgebras \ $U_{q_0}({\hat {sl_2}})\otimes U_{q_1}({\hat {sl_2}})$\ as in (\ref{algsine}) for $q_0=\exp(-2i\pi/\alpha_0^2)$, $q_1=\exp(-2i\pi/\alpha_1^2)$ and the identification :
\beqa
 {\cal Q}^{(a)}&=& {\cal Q}^{+}_{(a)}\ \ ; \  \  {\overline {\cal Q}}^{(a)} = {\overline {\cal Q}}^{-}_{(a)}\ ;\nonumber \\
{\cal Q}^{(a+2)}&=& {\cal Q}^{-}_{(a)}\ \ ; \  \  {\overline {\cal Q}}^{(a+2)} = {\overline {\cal Q}}^{+}_{(a)}\ \ \ \mbox{for}\ \ a=0,1 .
\eeqa
In particular, the choice $\alpha_0^2=\alpha_1^2=1$ corresponds to the free fermionic point of the two models indexed $(a)$.\\
\item $MSG_{(4|4)}(\beta,\beta;i\beta)$\\
For these values of the parameters, the multisine potential \ $\frac{\lambda}{\pi}\int d^2z \mathrm{mus}_0(\Phi)$\ in the action (\ref{action}) corresponds to \ $A^{(1)}_3$\ (simply-laced) ATFT for imaginary coupling. It writes :
\beqa
\frac{\lambda}{\pi}\int d^2z \sum_{i=0}^{3}n_i\exp(i\beta\ r_i.\Phi)
\eeqa
where $n_i=1$\  for \ $i\in[0,...,3]$ denote the Kac labels and $r_1=[0,-1,-1],\ r_2=[1,1,0],\ r_3=[0,1,1],\ r_0=-\sum_{i=1}^{3}r_i =[-1,1,0]$ are the simple roots of the algebra normalized as $r_i^2=2$. Then, the equations for the existence of non-local conserved charges associated to these parameters are obtained. For a ``normalized'' basis, one general solution\,\footnote{The cases \ $\beta'_0=0$\ or \ $\alpha'_0=0$ \ or \ $\alpha'_1=0$, which correspond to non-local conserved currents subalgebras, are not considered.} is given by : \ \ $\beta'_0=i\beta$\ and \ $\alpha'_0 = \alpha'_1=\beta$. The different deformations corresponding to the algebra of these non-local conserved charges are given by eq. (\ref{qab}), where $C_{a,b}$ is nothing else than different elements of the extended Cartan matrix associated to \ $A^{(1)}_3$\ affine Lie algebra. Since it is known that the quantum envelopping algebra  $U_q(A^{(1)}_3)$ can be described in the Chevalley basis by :
\beqa
\left[ H_a, E_b\right] &=& C_{a,b}E_b , \\ 
\left[H_a, F_b\right] &=& - C_{a,b}F_b,\nonumber \\
\left[E_a, F_b\right] &=& \delta_{ab}\frac{q^{H_a}-q^{-H_a}}{q-q^{-1}},\nonumber \\
\mbox{with} &&  H_{a}={\cal T}^{(a)}\ \  \ \mbox{the Cartan subalgebra and}\ a\in[0,....3],
\eeqa
the following identification for $q$ given by eq. (\ref{qab}) results : 
\beqa
{\cal Q}^{(a)}&=&cE_{a}q^{H_{a}/2}\ \ \ \ \ \ \mbox{and}\ \ \ \ \ \ {\overline {\cal Q}}^{(a)}=cF_{a}q^{H_{a}/2}
\eeqa
with $c$ some constant.

At this point, few remarks are in order. The $A_{3}^{(1)}$ ATFT for imaginary coupling $i\beta$ possesses topological soliton solutions which are in one-to-one correspondence with the $3$ fundamental representations of $A_{3}^{(1)}$ \cite{16,Holl}. It is possible to choose a soliton configuration for which $\Phi(x\rightarrow\infty)=0$ is satisfied. After some calculations, the different topological terms can be expressed in terms of the Cartan subalgebra defined as $n$ topological charges (not linearly independent) :
\beqa
H_a = -\frac{\beta}{2\pi}\int_t dx r_a.\partial_x\Phi\ \ \ \mbox{for}\ \ \ \ a\in[0,...,3].\label{topocartan} 
\eeqa
In representation (\ref{rep}), the r.h.s. of eq. (\ref{mualg}) becomes :
\beqa
T^{(a)}=\frac{\lambda}{4i\pi}\big( 1-q^{2H_a}\big)\ \ \ \ \mbox{for}\ \ \ \ a\in[0,...,3].
\eeqa
The conclusive understanding is that the more general hidden symmetry obtained from multicomplex framework is exactly identified to the quantum envelopping algebra $U_q(A^{(1)}_3)$ where $q$ is given by (\ref{qab}).
\end{itemize}
\begin{itemize}
\item $MSG_{(4|4)}(0,0;\beta_0)$\\
The MSG model then reduces to the sinh-Gordon model for the coupling parameter $\beta_0$. The non-local conserved charges symmetries simply reduce to the quantum group algebra already seen for the $MSG_{(2|2)}$ model but with an imaginary coupling parameter.  
\end{itemize}

\vspace{2mm}
\centerline{\bf Remarks  on other cases}
\vspace{3mm}

The case $MSG_{(3|3)}(0;\beta_0)$ reduces to the Bullough-Dodd model for the coupling parameter $\beta_0$. It corresponds to the ATFT based on $A^{(2)}_2$ non-simply laced Lie algebra with generalized Cartan matrix :
\begin{gather*}
{A_{ab}}= 
\begin{pmatrix}
 2 & -1\\
-4 & 2\\
\end{pmatrix}\label{A12}
\end{gather*}
As already seen, one may proceed by analogy. Here, the algebra of non-local conserved charges (\ref{alg}) is identified to the quantum envelopping algebra $U_q(A_2^{(2)})$. From the expression of constraint equations, \ $U_q({\hat {sl_2}})$\ is the corresponding algebra and another subalgebra exists described by $U_{q^4}({\hat {sl_2}})$ for $q$ given by (\ref{qab}) with $\beta_0=i\beta$.\\

This section closes with few words on an other example, the QFT $MSG_{(5|6)}(\alpha_0,\alpha_1;\beta_0,\beta_1)$. This QFT is integrable for specific values of the parameters \cite{multisine2}. It possesses a large variety of possible restrictions already analysed \cite{Fateev1,multisine,multisine2}. As done for $n=3$, it can be checked from (\ref{algodd}) that subalgebras of non-local charges exist which describe known quantum envelopping algebras. This theory admits a dual representation related to the non-linear sigma model and to the massive Thirring model coupled with ATFT \cite{multisine2}.  

\section{Discussion}
This paper was concerned with the formulation of a large variety of quantum field theories depending on $n-1$ parameters in terms of the multicomplex algebra. This is done using multicomplex extension of vertex operators. The essential benefit of this framework is the generic construction of a set of {\it multicomplex} non-local conserved charges with properly defined ``normalized'' MC-vertex operators.

They satisfy a $q-$deformed algebra for several constraints on parameters. This is done for any value of $n$, the dimension of the multicomplex space and any kind of representation ($\hbox{\it I\hskip -2.pt M \hskip -7 pt I \hskip - 3.pt \CC}_n \subset \hbox{\it I\hskip -2.pt M \hskip -7 pt I \hskip - 3.pt \CC}_{m}$ specifically). Then, for a specific choice of representation, the multicomplex algebra relates MSG theories to known quantum field theories like, for examples, pure and deformed affine Toda models, parafermionic sine-Gordon, integrable deformations of non-linear sigma model, etc... For such known cases, perfect agreement is found with results obtained in other frameworks. Let us now discuss some points and consider their eventual implications.
\begin{itemize}
\item {\it Quantization of MSG theories and unitarity}

For imaginary values of the parameters $\{\beta_a\}$, MSG action is not real for real-valued fields. Clearly, at quantum level this is somewhat questionable. An analogous situation appears in ATFTs with coupling $\beta$. For real values of $\beta$, the structure of the mass spectrum is such that a direct construction of a set of purely $S-$matrices is possible \cite{brad}. For imaginary $\beta$ values, the situation is much more complicated : at real-valued fields, the classical potential has discrete vacua. Solitons exist which connect these vacua and breathers of zero topological charge. Although they do not take real values, they have real energy and coupling \cite{unitarity}. One way to quantize these theories uses the semi-classical approach, but to our knowledge, this scheme is not completly satisfactory \cite{non-unitarity}. The alternative way consists in using the quantum group symmetry $U_q({\hat{\cal G}})$ of ATFTs based on the affine Lie algebra  ${\hat{\cal G}}$. Since topological charges are generators of this algebra and transform non-trivially the solitons, it is generally admitted that solitons-solitons $S-$matrices are proportional to the $R-$matrices of the symmetry algebra. This has been succesfully carried out for various ATFTs. Although nogeneral proof exists that $S-$matrices for lowest mass zero topological charge breathers are identical to the conjectured $S-$matrices of the lightest mass particle in the real coupling case (for $\beta\rightarrow i\beta$), it is found to be so in all cases. However, the $S-$matrices for the fundamental solitons which result from the QG method are not expected to be unitary (except for the SG model). Indeed, the issue of unitarity can be addressed directly at the level of the $S-$matrix, using QG or RSOS restriction \cite{QGR}. Since for $\{\beta_a\}\in {{i\mathbb R}^{*n/2-2}}$ the Hamitonian of MSG theories is complex, the theory may be expected to be non-unitary. For integrable cases, the two methods mentionned may be applied but as we saw, the quantum group structure is strongly related to specific contraints on parameters. 

\item {\it Integrability of MSG and $S-$matrix}

In ATFTs, the (infinite-dimensional) quantum symmetry $U_q({\hat{\cal G}})$ implies quantum integrability. It can be proved through the construction of an infinite number of commuting higher-spin conserved charges, provided by the infinitly many Casimir operators of $U_q({\hat{\cal G}})$. In fact, this approach is an alternative way to establish quantum integrability in ATFTs, as compared to the standard construction of {\it local} conserved charges. This later approach was retained in \cite{multisine}, but it is natural to attempt the first one in MSG theories. Due to the simple algebraic structure of multicomplex {\it non-local} conserved charges for $({\{\beta_a\}\in i\mathbb R^{*n/2-2}})$ described by eqs. (\ref{Qalg}) and (\ref{valtopo}), one may be able, for specific parameters related to eq. (\ref{Cab}), to extend these ideas to MSG theories. In particular, for the existence of fundamental multicomplex solitons fields, coproduct rules and multicomplex $R-$matrices, it should be the ideal basis for the construction of solutions of the Yang-Baxter equations associated to the MSG theories for $({\{\beta_a\}\in i\mathbb R^{*n/2-2}})$ and eq. (\ref{Cab}) satisfied. 
To analyse these problems, multicomplex extensions of $sl_2({\mathbb C})$ will be considered in a forthcomming publication.

\item {\it Subalgebras and restrictions of $MSG$}

From constraint equations (see appendix A, eqs. (\ref{cou})), specific values for $(\{\alpha'_a\},\{\beta'_a\})$ do lead to well-identified subalgebras.

Firstly, with $\beta'_a=0$ for all $a$, constraint equations (\ref{cou}) associated to ``normalized'' MC-vertex operators reduce to the standard equations associated to non-local conserved charges of the sine-Gordon model with solutions : $\alpha'_a=-\frac{2}{\alpha_a}$. Then, each $MSG(n|m)$ model in representation (\ref{rep}) possesses $\otimes_{a=0}^{E(n/2-1)}U_{q_a}({\hat {sl_2}})$ with $q_a=\exp(-2i\pi/\alpha^2_{a})$ as quantum envelopping subalgebras.

Consider for instance the case $MSG(4|4)(\alpha_0,\alpha_1;\beta_0)$. Using these subalgebras, a quantum group (QG) restriction is feasible with respect to the symmetry group $U_q({\hat {sl_2}})$ with $q=e^{-2i\pi/\alpha_0^2}$. The corresponding QFT gives the parafermionic sine-Gordon models \cite{frac,exp}.

Secondly, it follows from eqs.  (\ref{cou}), (\ref{Cklnorm}) and  (\ref{78}) in appendix A that $\{\alpha'_a\}=0$ can be choosen as a solution for all $a$. Any multicomplex identifies to its conjugate, hence only $n/2$ currents remain and $n/2$ currents are identical. Furthermore, the space of parameters is no longer restricted to $(\{\beta_a\}\in i{\mathbb R}^{*n/2-2})$, a consequence of eqs. (\ref{78}).

Although many different types of restrictions may be considered \cite{Fateev1}, their studies are beyond the scope of this paper.

\item {\it Restriction to $\{\alpha_a\}\in{\mathbb R}^{*n/2-1}$ and $\{\beta_a\}\in{\mathbb R}^{*n/2-2}$}

As explained in appendix D, for specific values of $n$, conserved currents can be found. In particular, for $n=5$ it is worth noting that the resulting charges possess a form very similar to those appearing in \cite{dual}. The model studied there presents a fermion-boson duality property. Non-local conserved charges exchange fermion and boson with the coupling. Hence, it would be interesting to investigate an hypothetic relation between duality transformation and multicomplex conjugation.

The conclusive message is that apparently disconnected models can be organized in terms of multicomplex algebra. Depending on the ratios of the coupling parameters, these models lead to known integrable QFTs. A geometrical interpretation can be given. In the vicinity of the critical point, a multicomplex manifold $\cal M$ characterized by $\{\alpha_a\},\{\beta_a\}$ exists. Several embedded submanifolds correspond to integrable QFTs. It is then possible to identify dual algebras in ATFTs (${\hat{\cal G}} \leftrightarrow {\hat {\cal G}}^\vee$) as two different representations of multicomplex algebra. More generally, algebraic duality could be understood as a geometrical (``non-minimal'') deformation of the multicomplex space.

\end{itemize}

\vspace{3cm}

\newpage

\appendix
\centerline{\bf \large Appendix A}
\vspace{0.6cm}
\centerline{\bf Conservation of non-local currents for even $n$}
\vspace{0.3cm} 

Here, we explain how to prove the existence of non-local charges using the extension of OPE rules within the multicomplex formalism. If OPEs between $J_{{\eta'}^{(k)}}$ and the holomorphic part (denoted ${\cal H}(.)$) of a multicomplex vertex operator ${\cal H}\big(x^{(l)}\big)=\exp(\eta^{(l)}.\phi(w))$:
\beqa
J_{{\eta'}^{(k)}}(z)x^{(l)}(w) \sim (z-w)^{-C^{k,l}({\eta'},\eta)} e^{({\eta'}^{(k)}+{\eta}^{(l)}).\phi(w)} +...\label{op}
\eeqa
behave (see section 3.1) as :
\beqa
&&J_{{\eta'}^{(k)}}(z)x^{(k)}(w) \sim (z-w)^{-2} e^{({\eta'}^{(k)}+{\eta}^{(k)}).\phi(w)}+... ,\label{cou} \\
&&J_{{\eta'}^{(k)}}(z)x^{(l)}(w) \sim (z-w)^{p} e^{({\eta'}^{(k)}+{\eta}^{(l)}).\phi(w)}+...\ \ \ \ \ \ \mbox{for other cases},\nonumber
\eeqa
where $p$ in the representation (\ref{rep}) is a $n \times n$ diagonal matrix with only real (or zero) positive numbers as entries and $C^{k,l}({\eta'},\eta)$ is given in eq. (\ref{Ckl}), then non-local conserved currents $J_{{\eta'}^{(k)}}$ exist\,\footnote{Similar results are obtained for the antiholomorphic part.}. 

\begin{itemize}
\item {\it A solution for ``normalized'' (``dual'') MC-vertex operators}
\end{itemize}
If we want to find a basis of ``normalized'' MC-vertex operators $J_{{\eta'}^{(k)\vee}}$ (we set ${\eta'}^{(k)}\rightarrow {\eta'}^{(k)\vee}$ in OPE (\ref{op})), we have :
\beqa
J_{{\eta'}^{(k)\vee}}(z)x^{(l)}(w) \sim (z-w)^{-C^{k,l}({\eta'}^{\vee},\eta)} e^{({\eta'}^{(k)\vee}+{\eta}^{(l)}).\phi(w)} +...\label{op1}
\eeqa
Then, we obtain a system of constraints associated to the multicomplex weights :
\beqa
C^{k,k}({\eta'}^{\vee}, \eta)&=& \Big[\sum_{a=0}^{\frac{n}{2}-2}C_{a,a}({\eta'}^{\vee},\eta) \big(-P^2_{a+k}\big)\Big]\nonumber \\
&& \ \ \ \ \ \ \ \ \ \ \ \  + C_{n/2-1,n/2-1}({\eta'}^{\vee},\eta) \big(-P^2_{\frac{n}{2}-1+k}\big),\nonumber
\eeqa
\beqa
C^{k,l}({\eta'}^{\vee}, \eta)&=& \Big(\sum_{a=0}^{\frac{n}{2}-2} C_{n/2-1,a}({\eta'}^{\vee},\eta)\delta_{a+l,k-1\ mod(\frac{n}{2})}\Big)\big(-P^2_{\frac{n}{2}-1+k}\big)\nonumber \\
&+& \Big(\sum_{a=0}^{\frac{n}{2}-2}C_{a,n/2-1}({\eta'}^{\vee},\eta)\delta_{a+k,l-1  mod(\frac{n}{2})}\Big)\big(-P^2_{\frac{n}{2}-1+l}\big),\label{Cklnorm}
\eeqa
\beqa
 C^{k,k+n/2}({\eta'}^{\vee}, \eta)&=&C^{k,k}({\eta'}^{\vee}, \eta)_{|\alpha_a'\rightarrow-\alpha_a'}\ \ \ \mbox{for}\ \ \ a\in[0,...,n/2-1],\nonumber\\
 C^{k,l+n/2}({\eta'}^{\vee}, \eta)&=&C^{k,l}({\eta'}^{\vee}, \eta).\nonumber
\eeqa 
with :
\beqa
C_{a,a}({\eta'}^{\vee},\eta)&=&\frac{2\big(\alpha'_a \alpha_a -\beta'_a \beta_a \big)}{({\alpha'}^2_a-{\beta'}^2_a)},\nonumber \\
C_{n/2-1,n/2-1}({\eta'}^{\vee},\eta)&=& \frac{2\big(\alpha'_{\frac{n}{2}-1}\alpha_{\frac{n}{2}-1} - \sum_{a=0}^{\frac{n}{2}-2}\beta'_a \beta_a \frac{m_a^2}{m_{\frac{n}{2}-1}^2}\big)}{\big({\alpha'}^2_{\frac{n}{2}-1} - \sum_{a=0}^{\frac{n}{2}-2}{\beta'_a}^2 \frac{m_a^2}{m_{\frac{n}{2}-1}^2}\big)},\nonumber
\eeqa
\beqa
C_{a,n/2-1}({\eta'}^{\vee},\eta)&=& \frac{m_a}{m_{\frac{n}{2}-1}}\frac{2\beta'_a \beta_a}{({\alpha'}_a^2-{\beta'}_a^2)},\label{78} \\
C_{n/2-1,a}({\eta'}^{\vee},\eta)&=& \frac{m_a}{m_{\frac{n}{2}-1}}\frac{2\beta'_a \beta_a}{({\alpha'}_{n/2-1}^2-\sum_{a=0}^{n/2-2}\frac{m_a^2}{m_{\frac{n}{2}-1}^2}{\beta'}_a^2)}.\nonumber
\eeqa
For $(\{\alpha_a\},\{\beta_a\}) \neq (0,0)$, the trivial solutions $\{\alpha_a'\}=\{\alpha_a\}$ and $\{\beta_a'\}=\{\beta_a\}$ are only possible if $\{\alpha_a\}\in {\mathbb R}^{*n/2-1}$,  $\{\beta_a\}\in i{\mathbb R}^{*n/2-2}$ or $\{\alpha_a\}\in i{\mathbb R}^{*n/2-1}$,  $\{\beta_a\}\in {\mathbb R}^{*n/2-2}$. It is easily understood from the structure of $C^{k,l}(\eta^{\vee},\eta)$ given by eq. (\ref{Cklnorm}) compared to constraint equations (\ref{cou}).

Consequently, for $\{\alpha_a\}\in {\mathbb R}^{*n/2-1}$ and $\{\beta_a\}\in i{\mathbb R}^{*n/2-2}$ we can choose ${\eta'}^\vee=\eta^\vee$ as a solution iff parameters $\{\alpha_a\},\{\beta_a\}$ satisfy $C_{a,a}(\eta^\vee,\eta)=2$ for all $a\in[0,...,n/2-1]$ and :
\beqa
C_{a,a+n/2}(\eta^\vee,\eta)\leq 0 \nonumber \\
C_{n/2-1,n-1}(\eta^\vee,\eta)\leq 0\nonumber
\eeqa
\vspace{0.5cm}

\vspace{0.6cm}
\centerline{\bf \large Appendix B}
\vspace{0.6cm}

\centerline{\bf Non-local charges algebra for odd $n$ in the representation (\ref{rep})}
\vspace{0.3cm}
It is straightforward to obtain the results for any odd $n$ multicomplex space using the substitution (\ref{subst}) in analogy to the even $n$ case. We obtain for $(a,b)\in[0,..., \frac{n-1}{2}-1]$ :
\beqa
q^2_{a,a}&=&q^2_{a+(n+1)/2,a+(n+1)/2}= \exp\big(-i\pi\frac{4}{({\alpha}_a^2-{\beta}_a^2)} \big), \nonumber \\
\ \ \ \ \ \ q^2_{(n-1)/2,(n-1)/2}&=&\exp\big(i\pi\frac{1}{\sum_{a=0}^{(n-1)/2-1}\frac{m_a^2}{m_{\frac{n-1}{2}}^2}{\beta}_a^2} \big), \nonumber \\
\ \ \ \ \ \ q^2_{a,a+(n+1)/2}&=& \exp\big(i\pi\frac{4( {\alpha}^2_{a} + {\beta}^2_{a} )}{({\alpha}_a^2-{\beta}_a^2)^2}\big),\label{algodd}
\eeqa
\beqa
q^2_{(n-1)/2,a}&=&q^2_{(n-1)/2,a+n/2}= \exp\big(i\pi\frac{2{\beta}^2_{a}}{({\alpha}_a^2-{\beta}_a^2)}\frac{1}{(\sum_{a=0}^{(n-1)/2-1}\frac{m_a^2}{m_{\frac{n-1}{2}}^2}{\beta}_a^2)} \frac{m_{a}}{m_{\frac{n-1}{2}}}\big), \nonumber \\
\ \ \ \ \ \ q^2_{a,b}&=&1 \ \ \ \ \mbox{in other cases},\nonumber \\
\ \ \ \ \ \ q^2_{a,b}&=&q^2_{a,b+(n+1)/2}.\nonumber
\eeqa
Furthermore, the topological term for $a\in[0,...,\frac{n-1}{2}-1]$ is given by :
\beqa
T^{(a)}&=& \frac{\lambda}{in\pi} \int_t dx\partial_x \exp\Big(-i\big(\frac{-2{\alpha}_{a}}{({\alpha}_a^2-{\beta}_a^2)}+ {\alpha}_{a}\big).(\phi +{\overline \phi})_{a}\label{topo2odd} \\
&&\ \ \ \ \ \ \ \ \ \ \ \ \ \ \ \ \ \ +  \big(\frac{-2{\beta}_{a}}{({\alpha}_a^2-{\beta}_a^2)}+ {\beta}_{a}\big).(\varphi +{\overline \varphi})_{a}\Big)\nonumber
\eeqa
and
\beqa
T^{(\frac{n-1}{2})}&=&\frac{\lambda}{in\pi} \int_t dx\partial_x \exp\Big( - \ \sum_{a=0}^{(n-1)/2-1}\Big(\big(\frac{-2{\beta}_a}{\sum_{a=0}^{(n-1)/2-1}\frac{m_a^2}{m_{\frac{n-1}{2}}^2}{\beta}_a^2}+\beta_a\big)\frac{m_a}{m_{(n-1)/2}}\Big).(\varphi +{\overline \varphi})_{a}\Big).\nonumber
\eeqa
For $\{\alpha_a\}\in \mathbb{R}^{*n/2-1}$ and $\beta_a\in i\mathbb{R}^{*n/2-2}$, the topological terms get simplify easily as shown for the even $n$ case. Explicitely, for specific values of the parameters $(\{\alpha_a\},\{\beta_a\})$ which identify the corresponding MSG model to ATFT based on Lie algebra $G$, it can be expressed in terms of the Cartan subalgebra of $G^{\vee}$, the dual affine Lie algebra of $G$.  

\vspace{0.5cm}

\vspace{0.6cm}
\centerline{\bf \large Appendix C}
\vspace{0.6cm}

From eqs. (\ref{Cklnorm}), it is possible to define a ``deformed'' version similar to the standard Cartan matrix where the deformation is characterized by parameters $(\{\alpha_a\}\in {\mathbb R}^{*n/2-1},\{\beta_a\}\in i{\mathbb R}^{*n/2-2})$. The Cartan matrix of the affine Lie algebra $\hat{\cal G}$ is defined as :
\beqa
A_{ab}=\frac{2r_a.r_b}{r_a^2} \in {\mathbb Z}
\eeqa
where $r_a,r_b$ are simple roots associated to the affine Lie algebra $\hat{\cal G}$. We similarly define the ``deformed'' Cartan matrix by :

\beqa
C^{k,l}(\eta^\vee,\eta)=\frac{2\eta^{(k)}.\eta^{(l)}}{(\eta^{(k)})^2}\label{80}
\eeqa
in representation (\ref{rep}), it gives for $(a,b)\in[0,...,n/2-2]$ :
\beqa
C_{a,a}(\eta^\vee,\eta)&=& \Big(\pi\big[C^{n/2-a,n/2-a}(\eta^{\vee},\eta)\big]\Big)_{00}= 2,\nonumber \\ 
C_{a,a+n/2}(\eta^\vee,\eta)&=& \Big(\pi\big[C^{n/2-a,n-a}(\eta^{\vee},\eta)\big]\Big)_{00}= -\frac{2(\alpha_a^2+\beta_a^2)}{\alpha_a^2-\beta_a^2},\label{Cab} \\
 &=&C_{a+n/2,a}(\eta^\vee,\eta),\nonumber \\
C_{n/2-1,n-1}(\eta^\vee,\eta)&=&\Big(\pi\big[C^{1,n/2+1}(\eta^{\vee},\eta)\big]\Big)_{00}= -\frac{2\big(\alpha^2_{\frac{n}{2}-1} + \sum_{a=0}^{\frac{n}{2}-2}\beta^2_a \frac{m_a^2}{m_{\frac{n}{2}-1}^2}\big)}{\big(\alpha^2_{\frac{n}{2}-1} - \sum_{a=0}^{\frac{n}{2}-2}\beta^2_a \frac{m_a^2}{m_{\frac{n}{2}-1}^2}\big)},\nonumber \\ 
&=& C_{n-1,n/2-1}(\eta^\vee,\eta),\nonumber \\
C_{a,n/2-1}(\eta^\vee,\eta)&=&\Big(\pi\big[C^{n/2-a,1}(\eta^{\vee},\eta)\big]\Big)_{00}=\frac{2\beta_a^2}{\alpha_a^2-\beta_a^2}\frac{m_a}{m_{\frac{n}{2}-1}},\nonumber \\ 
C_{n/2-1,a}(\eta^\vee,\eta)&=&\Big(\pi\big[C^{1,n/2-a}(\eta^{\vee},\eta)\big]\Big)_{00}=\frac{2\beta_a^2}{\big(\alpha^2_{\frac{n}{2}-1} - \sum_{a=0}^{\frac{n}{2}-2}\beta^2_a \frac{m_a^2}{m_{\frac{n}{2}-1}^2}\big)}\frac{m_a}{m_{\frac{n}{2}-1}},\nonumber \\
C_{a,b}(\eta^\vee,\eta)&=&0 \ \ \ \ \ \ \ \ \mbox{in other case}.\nonumber\\
\mbox{and}\ \ \ \ C_{a+n/2,b}(\eta^\vee,\eta)&=&C_{a,b}(\eta^\vee,\eta).
\eeqa
As for the standard Cartan matrix, we affect a node indexed $(a)$ to each value of $a\in[0,...,n/2-1]$.  However, for $\alpha_{n/2-1}=0$, $C_{n/2-1,n-1}(\eta^\vee,\eta)=C_{n/2-1,n/2-1}(\eta^\vee,\eta)=2$, so one sees that conserved currents ${\cal J}^{(n/2-1)}={\cal J}^{(n-1)}$. Similarly, the two nodes $((n-1)/2)$ and $(n)$ are identical for odd $n$. This can be generalized as follows : each time we have $\alpha_a=0$, the two nodes $(a)$ and $(a+n/2)$ collapse together. This results from the multicomplex symmetry between $P_a \rightarrow P_{a+n/2} =-P_a$, relating one node $(a)$ and its conjugate $(a+n/2)$, which disappears for $\alpha_a=0$. 

A crucial condition to obtain a quantum group symmetry for (\ref{Qalg}) corresponds to $C_{a,b}(\eta^\vee,\eta)\in {\mathbb Z}_{-}$ for $(a\neq b)\in[0,...,n/2-1]$. We may ask for which values of parameters it can be satisfied and then, how  the resulting Dynkin diagrams look like. We have studied the cases $n=3$ and $n=4$.

\begin{itemize}

\item{Multicomplex dimension $n=3$}

For eq. (\ref{80}) to be satisfied, we choose \ $\beta_0^2=\frac{N-2}{N+2}\alpha_0^2$,\  then :
\beqa
C_{0,2}(\eta^\vee,\eta)=-N,\ \ \ \ \ \ C_{0,1}(\eta^\vee,\eta)=C_{2,1}(\eta^\vee,\eta)=\frac{m_0}{m_1}(N-2)\in {\mathbb Z}_{-}, \\ 
C_{1,0}(\eta^\vee,\eta)=C_{1,2}(\eta^\vee,\eta)=-\frac{m_1}{m_0}\in {\mathbb Z}_{-}.\nonumber
\eeqa
- For $N=0$, $\beta_0^2=-\alpha_0^2$ : we must take $\frac{m_0}{m_1}=1$ $(MSG(3|3))$ or $\frac{m_0}{m_1}=\frac{1}{2}$ $(MSG(3|4))$.\\

- For $N=1$, $\beta_0^2=-\frac{1}{3}\alpha_0^2$ : we must take $\frac{m_0}{m_1}=1$ $(MSG(3|3))$.

\item{Multicomplex dimension $n=4$}

Similarly, we choose \ $\beta_0^2=\frac{N-2}{N+2}\alpha_0^2$ \  and \ $\beta_1^2=\frac{M-2}{M+2}\alpha_1^2\frac{m_1^2}{m_0^2}$. It gives : 
\beqa
&&C_{0,2}(\eta^\vee,\eta)=-N, \\
&&C_{0,1}(\eta^\vee,\eta)=C_{2,1}(\eta^\vee,\eta)=C_{0,3}(\eta^\vee,\eta)=C_{2,3}(\eta^\vee,\eta)=\frac{m_0}{2m_1}(N-2)\in {\mathbb Z}_{-},\nonumber \\
&&C_{1,3}(\eta^\vee,\eta)=-M,\nonumber \\
&&C_{1,0}(\eta^\vee,\eta)=C_{1,2}(\eta^\vee,\eta)=C_{3,0}(\eta^\vee,\eta)=C_{3,2}(\eta^\vee,\eta)=\frac{m_1}{2m_0}(M-2)\in {\mathbb Z}_{-},\nonumber
\eeqa
The only possibility is $N=M=0$ which gives $\beta_0^2=-\alpha_0^2$ and $\alpha_0^2=\alpha_1^2$ for $\frac{m_0^2}{m_1^2}=1$ $(MSG(4|4))$. From eq. (\ref{Qalg}) in representation (\ref{rep}) compared to convention (\ref{qab}), we have $C_{a,b}=C_{a,b}(\eta,\eta^\vee)=C_{b,a}(\eta^\vee,\eta)$. For each hidden symmetry, we have listed in figure 2 the Dynkin diagrams associated respectively to $C_2^{(1)}$, $B_2^{(2)}=D_3^{(2)}$, $A_2^{(1)}$ and $A_3^{(1)}$ affine Lie algebras.

\end{itemize}

\vspace{5mm}

\centerline{\epsfig{file=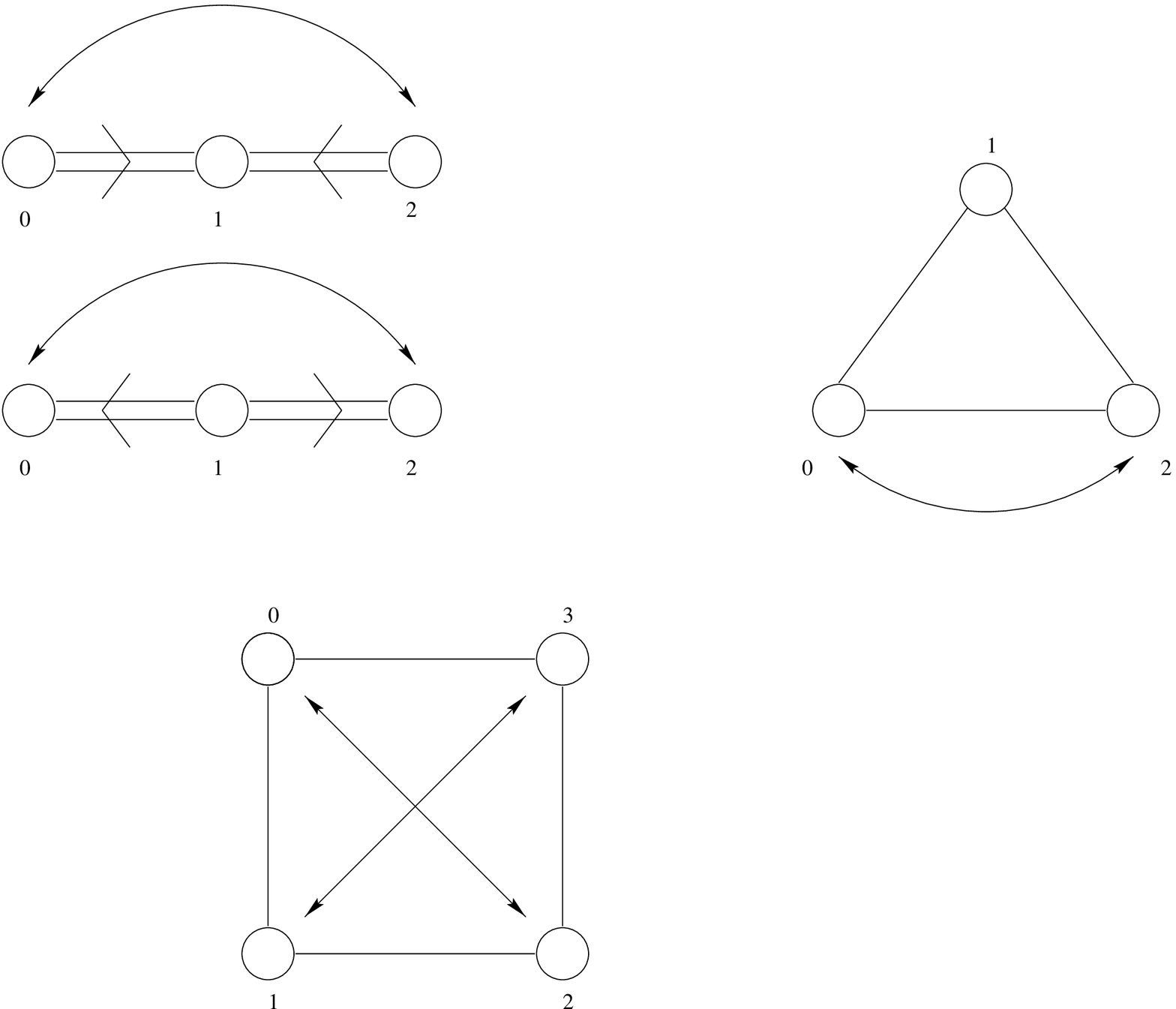,height=95mm,width=95mm}}
\vspace{5mm}

\centerline{\small Figure 2 : Arrows symbolise the exchange under multicomplex}
\centerline{\small conjugation $(a)\rightarrow(a+n/2)$. The number of links is $|C_{a,b}(\eta,\eta^\vee)||C_{a,b}(\eta^\vee,\eta)|$}
\centerline{\small where big arrows goes from $a$ to $b$ if $|C_{a,b}(\eta,\eta^\vee)|>|C_{a,b}(\eta^\vee,\eta)|$.}

\newpage

\centerline{\bf \large Appendix D}
\vspace{0.6cm}

\centerline{\bf Deformed algebra associated to``unormalized'' solutions}
\vspace{2mm}

\begin{itemize}
\item {\it A solution for ``unormalized'' MC-vertex operators} ($n=4,5$)
\end{itemize}
Since the different generators of the multicomplex algebra verify (\ref{pa}) and are linearly independent, we have to find a solution - associated to the existence of a non-local conserved currents parametrized by $(\{{\alpha'}_a\},\{{\beta'}_a)\}$ - to the following system of constraints (using (\ref{Ckl}) for ``unormalized'' vertex operators). For even $n$ :
\beqa
&&\Big(-\alpha'_a \alpha_a + \beta'_a \beta_a\Big) = 2 ,\ \ \ \ \ \ \ \ \ \Big(-\alpha'_ {\frac{n}{2}-1}\alpha_{\frac{n}{2}-1} + \sum_{a=0}^{\frac{n}{2}-2}\beta'_a \beta_a \frac{m_a^2}{m_{\frac{n}{2}-1}^2}\Big) = 2, \label{cont3} \\
&&\Big(\alpha'_a \alpha_a + \beta'_a \beta_a\Big) \in {\mathbb Z}_{-} ,\ \ \ \ \ \ \ \ \ \ \ \ \ \ \Big(\alpha'_ {\frac{n}{2}-1}\alpha_{\frac{n}{2}-1} + \sum_{a=0}^{\frac{n}{2}-2}\beta'_a \beta_a \frac{m_a^2}{m_{\frac{n}{2}-1}^2}\Big) \in {\mathbb Z}_{-},  \nonumber \\
&&\mbox{and}\ \ \ \ \ \ \frac{m_a}{m_{\frac{n}{2}-1}}\beta'_a \beta_a  \in {\mathbb Z}_{+}^{*} \ \ \ \ \ \ \ \ \ \ \ \ \ \ \ \ \ \ \ \ \ \mbox{for all}\ \ a\in[0,...,n/2-2].\nonumber
\eeqa
For instance, one obvious solution is given by : 
\beqa
&&\alpha_a'=-1/\alpha_a,\ \ \ \ \beta_a'=1/\beta_a\ \ \ \mbox{for}\ \ \ a\in[0,...,n/2-2],\label{solun} \\
&&\alpha'_{n/2-1}= -\frac{\big(2-\sum_{a=0}^{n/2-2}\frac{m^2_a}{m^2_{n/2-1}} \big)}{\alpha_{n/2-1}}\nonumber
\eeqa
iff
\beqa
\sum_{a=0}^{n/2-2}\frac{m_a^2}{m^2_{n/2-1}}= 1.\label{rest}
\eeqa
However, as $\frac{m_a}{m_{n/2-1}}\in{\mathbb Z}_{+}^{*}$, it is only possible for $n=4$ in the ``minimal'' representation $(\frac{m_0}{m_1}=1)$. For odd $n$, we set (\ref{subst}). It gives the solutions : 
\beqa
&&\alpha_a'=-1/\alpha_a,\ \ \ \ \beta_a'=1/\beta_a\ \ \ \mbox{for}\ \ \ a\in[0,...,n/2-2]\label{soluodd}
\eeqa
but instead of (\ref{rest}), we must have : 
\beqa
\sum_{a=0}^{(n-3)/2}\frac{m_a^2}{m^2_{(n-1)/2}}= \frac{1}{2} \ \ \ \mbox{with}\ \ \ \frac{2m_a}{m_{(n-1)/2}}\in{\mathbb Z}^{*}_{+}\label{restodd}
\eeqa
which can be only satisfied for $n=5$ iff $\frac{m_0}{m_2}=\frac{m_1}{m_2}=\frac{1}{2}$.

Then, for $n=4,5$, we may choose ``unormalized'' MC-vertex operators as a convenient basis of hypothetical non-local conserved currents : instead of the ``normalized'' basis, we are not forced here to restrict the space of parameters. In fact, whereas a ``normalized'' basis for $(\{\alpha_a\}\in{\mathbb R}^{*n/2-1},\{\beta_a\}\in {\mathbb R}^{*n/2-2})$ does not exist (as we see comparing eqs. (\ref{Cklnorm}), (\ref{cou}) and (\ref{op1})), an ``unormalized'' one exists. We detail the algebraic structure below. However, the structure in the r.h.s. of eq. (\ref{mualg}) does not simplify as for $(\{\alpha_a\}\in{\mathbb R}^{n/2-1},\{\beta_a\}\in i{\mathbb R}^{n/2-2})$.  

If $(\{\alpha_a\},\{\beta_a\})$ satisfy eqs. (\ref{cont3}), the non-local currents $J_{{\eta'}^{(k)}}$ are conserved for all $k$. We obtain :
\beqa
{\cal J}^{(a)}&=&\exp\big(-i\alpha'_a\phi_a +\beta'_a\varphi_a\big)\ \ \ \ \ \mbox{for}\ \ a\in[0,...,n/2-2],\\
{\cal J}^{(n/2-1)}&=&\exp\big(-i{\alpha'}_{n/2-1}\phi_{n/2-1} - \sum_{a=0}^{n/2-2}\beta'_a \frac{m_a}{m_{n/2-1}}\varphi_{n/2-1}  \big).\nonumber
\eeqa
It is sufficient to perform the inverse of transformation (\ref{change}) in (\ref{alg}) and (\ref{topo2}) to obtain the corresponding quantities for ``unormalized'' MC-operators except for $T^{(n/2-1)}$. $T^{(n/2-1)}$ for the above solutions is given by :
\beqa
T^{(n/2-1)}&=&\frac{\lambda}{in\pi} \int_t dx\partial_x \exp\Big( -i\big(   \alpha'_{n/2-1}  + \alpha_{n/2-1}\big).(\phi +{\overline \phi})_{n/2-1}\label{topo3}\nonumber \\
&& - \ \sum_{a=0}^{n/2-2}\big(\beta'_a + \beta_a \big)\frac{m_a}{m_{n/2-1}}.(\varphi +{\overline \varphi})_{a}.
\eeqa

\paragraph*{Aknowledgements}
I am very grateful to V. A. Fateev, P. Grang\'e, A. Neveu, F. A. Smirnov, J. Thierry-Mieg and especially M. Rausch de Traubenberg and D. Reynaud for useful discussions.

\end{document}